\documentclass{ieeeaccess}
\usepackage{cite}
\usepackage{amsmath,amssymb,amsfonts}
\usepackage{algorithmic}
\usepackage{graphicx}
\usepackage{textcomp}
\usepackage{booktabs}
\usepackage{multirow}
\usepackage{graphicx}
\usepackage[font={sf,scriptsize},
            labelfont={bf}, 
            caption=false]
           {subfig}
\usepackage{bm}
\makeatletter
\AtBeginDocument{\DeclareMathVersion{bold}
\SetSymbolFont{operators}{bold}{T1}{times}{b}{n}
\SetSymbolFont{NewLetters}{bold}{T1}{times}{b}{it}
\SetMathAlphabet{\mathrm}{bold}{T1}{times}{b}{n}
\SetMathAlphabet{\mathit}{bold}{T1}{times}{b}{it}
\SetMathAlphabet{\mathbf}{bold}{T1}{times}{b}{n}
\SetMathAlphabet{\mathtt}{bold}{OT1}{pcr}{b}{n}
\SetSymbolFont{symbols}{bold}{OMS}{cmsy}{b}{n}
\renewcommand\boldmath{\@nomath\boldmath\mathversion{bold}}}
\makeatother

\def\BibTeX{{\rm B\kern-.05em{\sc i\kern-.025em b}\kern-.08em
    T\kern-.1667em\lower.7ex\hbox{E}\kern-.125emX}}

\begin{document}
\history{Date of publication xxxx 00, 0000, date of current version xxxx 00, 0000.}
\doi{10.1109/ACCESS.2024.0429000}

\title{Parametric Digital Twins for Preserving Historic Buildings: A Case Study at Löfstad Castle in Östergötland, Sweden}
\author{\uppercase{Zhongjun Ni}\authorrefmark{1},
\uppercase{Jelrik Hupkes}\authorrefmark{2},
\uppercase{Petra Eriksson}\authorrefmark{2},
\uppercase{Gustaf Leijonhufvud}\authorrefmark{2},
\uppercase{Magnus Karlsson}\authorrefmark{1}, and \uppercase{Shaofang Gong}\authorrefmark{1}}

\address[1]{Department of Science and Technology, Linköping University, Campus Norrköping, 60174 Norrköping, Sweden}
\address[2]{Department of Art History, Uppsala University, Campus Gotland, 62157 Visby, Sweden}
\tfootnote{This study has been financially supported in part by 
the Swedish Innovation Agency (Vinnova) and in part by the Swedish Energy Agency (Energimyndigheten).}

\markboth
{Author \headeretal: Preparation of Papers for IEEE TRANSACTIONS and JOURNALS}
{Author \headeretal: Preparation of Papers for IEEE TRANSACTIONS and JOURNALS}

\corresp{Corresponding author: Zhongjun Ni (e-mail: zhongjun.ni@liu.se).}

\begin{abstract}
This study showcases the digitalization of Löfstad Castle in Sweden to contribute to preserving its heritage values. The castle and its collections are deteriorating due to an inappropriate indoor climate. To address this, thirteen cloud-connected sensor boxes, equipped with 84 sensors, were installed throughout the main building, from the basement to the attic, to continuously monitor various indoor environmental parameters. The collected extensive multi-parametric data form the basis for creating a parametric digital twin of the building. The digital twin and detailed data analytics offer a deeper understanding of indoor climate and guide the adoption of appropriate heating and ventilation strategies. The results revealed the need to address high humidity problems in the basement and on the ground floor, such as installing vapor barriers. Opportunities for adopting energy-efficient heating and ventilation strategies on the upper floors were also highlighted. The digitalization solution and findings are not only applicable to Löfstad Castle but also provide valuable guidance for the conservation of other historic buildings facing similar challenges.
\end{abstract}


\titlepgskip=-21pt

\maketitle

\section{Introduction}
\label{sec:introduction}

\PARstart{D}{igital} transformation of the built environment brings opportunities to improve efficiency and sustainability for building operations and maintenance~\cite{jia_adopting_2019}. For example, real-time data visualization allows facility managers to monitor building conditions continuously and take timely actions~\cite{ni_sensing_2021}. This transformation relies on integrating digital technologies such as Internet of Things (IoT), cloud and edge computing, as well as reality capture technologies like 3D laser scanning. IoT enables continuous data collection, helping facility managers maintain up-to-date knowledge bases~\cite{atta_digital_2020}. Cloud computing offers scalable resources for data processing~\cite{liu_methodology_2021}, while edge computing addresses latency and privacy concerns by moving computation closer to the network edge~\cite{hua_edge_2023}. Additionally, 3D laser scanning is increasingly used for modeling existing buildings, making it ideal for restoration projects~\cite{yang_review_2020}.

As data accumulate, analytics becomes crucial for optimizing building operations and maintenance. A key aspect of building operations is indoor climate control. It can benefit from data-driven approaches such as load forecasting~\cite{ni_study_2024}, model-based predictive controls~\cite{drgoňa_all_2020}, and occupant-centric controls~\cite{gunay_data_2019, hammar_realestatecore_2019}. Furthermore, real-time data combined with data analytics can be used to create a digital twin of a building. The digital twin is a comprehensive representation of the building that includes all necessary information relevant to its lifecycle~\cite{boschert_digital_2016}. A digital twin can support decision-making across various building operations by assessing current conditions~\cite{zhang_automatic_2022}, diagnosing problems~\cite{ni_enabling_2022}, and predicting future status~\cite{ni_edge_2024}.

Conservation of historic buildings is a vital scenario for digital transformation of the built environment. Unlike ordinary buildings, where human comfort is usually the primary focus, building conservation must also prioritize the protection of the building itself and housed collections~\cite{leijonhufvud_standardizing_2018}. The conservation process typically involves diagnosing deterioration, conducting necessary interventions and treatments, and implementing preventive measures~\cite{doerr_ontologies_2009}. Digital twins can contribute to this process by enabling continuous monitoring, in-depth analysis, and proactive action. Monitoring involves collecting data to update the digital representation of the building, serving as a diagnostic tool for conservation~\cite{balen_preventive_2015}. Analyzing these data is essential for identifying and addressing potential problems before they escalate, ensuring long-term building conservation. The action phase includes retrofitting and optimizing control measures to mitigate damage. In addition to these aspects, energy efficiency is critical in preserving historic buildings, aiming to reduce energy use, improve human comfort, and preserve heritage values~\cite{eriksson_balancing_2019, ni_improving_2021}.

Previous studies have explored the creation of digital twins to preserve historic buildings~\cite{angjeliu_development_2020, zhang_automatic_2022, ni_enabling_2022}. A key challenge in this process is organizing heterogeneous data from diverse sources into a unified system~\cite{balaji_brick_2018}. This diversity leads to fragmented data and inconsistent information flows across different buildings, system and equipment vendors, and locations. To improve interoperability within subsystems of a building and promote broader adoption across various buildings, it is crucial to adopt a common data standard for digital twin solutions. A semantic approach using ontologies can help address data heterogeneity and enhance interoperability~\cite{gyrard_building_2018}. While some studies, such as~\cite{ni_enabling_2022}, have combined IoT and ontology to create digital twins for historic buildings, these solutions are often tied to a specific cloud service, i.e., there are no alternatives from other cloud vendors, limiting their adaptability and wider use across different cloud environments.

This study aims to contribute to preserving historic buildings by creating digital twins that extend beyond the traditional 3D models commonly used in building information modeling (BIM). Conventional BIM approaches mainly handle static attributes. This work employed parametric models to capture both essential static attributes and dynamic behaviors of buildings, which enhances more comprehensive environmental monitoring and analysis. The main contributions of this work are:
\begin{itemize}
\item Presented a cloud-based solution to create parametric digital twins for conservation of historic buildings by
combining graphical and relational data models. The graphical model stores contextual data, such as spatial information of rooms and sensor locations. In contrast, the relational model captures time series data like indoor climate and outdoor weather conditions. This integration facilitates real-time monitoring and provides data-driven insights into building operations. A full implementation of this approach was also given. The implementation employed and extended the Brick ontology~\cite{balaji_brick_2018} to ensure data consistency and interoperability, enhancing the integration of diverse data sources and applications. To avoid reinvention, several public cloud services provided by Microsoft Azure~\cite{azure_products_online} were utilized to build the cloud part of the solution.
\item Conducted a comprehensive case study in a historic building in Östergötland, Sweden. Thirteen cloud-connected sensor boxes, equipped with 84 sensors, were deployed across all floors. This extensive sensor distribution enabled the creation of a parametric digital twin that not only captured environmental conditions but also supported advanced data analytics to assess the indoor climate. The analysis revealed the necessity of implementing measures to address high humidity problems in the basement and on the ground floor, e.g., installing vapor barriers. Meanwhile, energy-efficient heating and ventilation strategies were identified for the upper floors. Notably, the presented solution and insights derived from this study are transferable to other historic buildings facing similar problems, providing a flexible framework that can be adapted to various environments and conservation challenges.
\end{itemize}

The remainder of this paper is structured as follows. After discussing related work in Section~\ref{sec:related_work}, the detailed methodology for creating parametric digital twins is described in Section~\ref{sec:methodology}. Then, a case study, including a detailed description of the building, deployment of local devices, developed applications, and methods to evaluate indoor climate, is given in Section~\ref{sec:case_study}. After that, Section~\ref{sec:results_and_discussion} presents and discusses the obtained results. The last section concludes the paper.

\section{Related Work}
\label{sec:related_work}

This section begins by reviewing previous research on indoor climate in historic buildings, focusing on evaluating indoor climate, factors influencing indoor climate, and measures to improve indoor climate. Then, digital twin applications in buildings, including historic buildings, are summarized, covering the creation of digital twins, data models, and specific application cases.

\subsection{Indoor Climate in Historic Buildings}

Maintaining a suitable indoor climate is crucial for preserving historic buildings and housed collections~\cite{leijonhufvud_standardizing_2018}. Two key indoor environmental parameters are temperature and relative humidity (RH). Drastic short-term fluctuations in these two parameters can damage cultural artifacts, especially those made of organic hygroscopic materials~\cite{en_15757_conservation_2010}. Several standards have been developed to determine appropriate temperature and RH. For instance, the European standard EN 15757:2010~\cite{en_15757_conservation_2010} provides guidelines for specifying target ranges of temperature and RH to prevent material deterioration caused by strain-stress cycles. Unlike traditional single target values~\cite{gt_museum_1986}, EN 15757:2010 proposes flexible target ranges for these parameters. When a historic building suffers from high RH problems, it is essential to identify moisture sources. The European standard EN 16242:2012~\cite{en_16242_conservation_2012} recommends using the humidity mixing ratio (MR)--the mass of water vapor divided by the mass of dry air--as a diagnostic tool. A comparison between indoor and outdoor MRs can indicate a moisture source or sink in a building~\cite{brostrom_climate_2015}. Studies have also examined the risk of mold growth due to inappropriate temperature and RH. The isopleth system for substrate category I~\cite{sedlbauer_prediction_2001} is commonly used to assess mold risk, covering all molds typically found in buildings.

Indoor climate in historic buildings can be affected by factors like air exchange, human activities, and operation of heating, ventilation, and air conditioning (HVAC) systems. Lowering the air exchange rate can stabilize indoor climate, thus reducing the risk of mechanical damage to hygroscopic materials~\cite{luciani_influence_2013}. Besides, for historic buildings with massive walls, the walls can act as thermal buffers to moderate fluctuations in temperature and RH~\cite{varas_muriel_fluctuations_2014, pisello_coupling_2018}. Occupants also affect indoor climate. Many studies~\cite{ferdyn_grygierek_monitoring_2016, assimakopoulos_comparison_2017, pisello_coupling_2018, caro_evaluation_2020, uring_assessment_2020, nawalany_analysis_2021, silva_impact_2021, vázquez_torres_impact_2022, ni_enabling_2022} have investigated the impact of occupancy levels on temperature, RH, and carbon dioxide (CO\textsubscript{2}) concentration. Higher occupancy levels usually lead to higher rises in temperature, while changes in RH result from combined effects of heat and moisture gained from occupants~\cite{camuffo_impact_2004}. The presence of occupants raises CO\textsubscript{2} concentration more quickly than temperature. Owing to this characteristic, studies~\cite{caro_evaluation_2020, ni_enabling_2022} have used indoor CO\textsubscript{2} concentration to indicate occupancy level. Notably, a continuous high occupancy level can cause CO\textsubscript{2} concentration to exceed recommended levels, and natural ventilation may not be sufficient to maintain good indoor air quality~\cite{yüksel_review_2021}. Since historic buildings were typically not built with HVAC systems, adding these systems without careful consideration could harm building conservation~\cite{bay_assessment_2022}, e.g., disrupting historical climate and potentially damaging cultural artifacts~\cite{sahin_investigation_2017}. Even in refurbished historic buildings equipped with HVAC systems, improper scheduling can lead to problems such as stuffy air and unsuitable temperatures~\cite{bakhtiari_evaluation_2020}.

Research on improving indoor climate in historic buildings has employed digitalization approaches like real-time environmental monitoring~\cite{zhang_culturebee_2011, ni_sensing_2021, barsocchi_wireless_2021, colace_iot_2021, silva_impact_2021} and computer simulation~\cite{schellen_overview_2007, schijndel_application_2008, bay_assessment_2022}. Real-time monitoring enables timely maintenance and intervention~\cite{colace_iot_2021} through continuously collecting data about indoor climate. In a historic building where installing mechanical ventilation was impractical, Silva and Henriques~\cite{silva_impact_2021} proposed restricting the number of visitors by monitoring CO\textsubscript{2} concentration. When indoor CO\textsubscript{2} concentration exceeded a certain threshold compared to outdoor level, entry was restricted until the difference decreased under the threshold~\cite{silva_impact_2021}. Computer simulation involves creating models based on laboratory and on-site measurements. These models help understand physical processes and recommend improvements in indoor climate control~\cite{schellen_overview_2007}. Bay~\textit{et al.}~\cite{bay_assessment_2022} used a computational fluid dynamics simulation model to identify effective natural ventilation scenarios for a historic building in a hot and humid climate. Schijndel~\textit{et al.}~\cite{schijndel_application_2008} developed an integrated heat, air, and moisture model to design HVAC systems for a museum. The results suggested keeping old wallpaper fragments in a showcase and adopting a new control strategy to meet environmental requirements~\cite{schijndel_application_2008}. Some researchers~\cite{leijonhufvud_standardizing_2018} also advocated new standards to provide practical guidelines that can be integrated into existing management processes, rather than focusing solely on outcomes.

\subsection{Digital Twins for Buildings}

A digital twin of a building has three essential parts: a model of the building, an evolving set of data relating to the building, and a method for dynamically updating the model with the data~\cite{wright_how_2020}. Models used in previous studies can be classified into three categories: 3D geometric models, information models, and parametric models. A 3D geometric model reconstructs the architecture of a building using techniques like 3D computer graphics, photogrammetry, and laser scanning~\cite{yang_review_2020}. An information model creates a database containing attributes of building components, such as materials, connections, and 3D geometry~\cite{yang_review_2020}. A parametric model focuses on defining and describing key attributes and behaviors of a building~\cite{ni_enabling_2022}. While 3D geometric and information models primarily document static aspects of buildings~\cite{delgado_digital_2021} and are helpful in managing and restoring historic buildings~\cite{yang_review_2020}, parametric models leverage ongoing data updates for data analytics and preventive maintenance~\cite{luo_overview_2021}.

Integrating diverse information into parametric models requires independent and organized methods~\cite{bruno_historic_2018}. Data related to historic buildings are heterogeneous and multidisciplinary, requiring expertise in various fields such as information technology, history, and architecture~\cite{noor_modeling_2019}. Therefore, it is necessary to describe these data consistently to ensure interoperability between tools and applications~\cite{luo_overview_2021}. 
Ontologies play an essential role in organizing related data objects and defining their relationships, which aids in the linking and composition of concepts~\cite{luo_overview_2021}.
Ontologies are especially practical for representing data about historic buildings due to their graphical structure and ability to self-describe entities and their connections~\cite{noor_modeling_2019}. Two notable ontologies in the domain of digital transformation of buildings are
RealEstateCore~\cite{hammar_realestatecore_2019} and Brick~\cite{balaji_brick_2018}.
These two ontologies have cooperated for years to create alignments that make it easier to translate between the two schemas.

Several studies have explored creating digital twins for buildings by integrating various technologies to enhance functionality and efficiency. Khajavi~\textit{et al.}~\cite{khajavi_digital_2019} created a digital twin of a building facade using a wireless sensor network to model ambient environment and included a tool for visualizing lighting data. Similarly, Rosati~\textit{et al.}~\cite{rosati_air_2020} combined IoT sensor networks with BIM to automatically update indoor air quality readings in BIM models, reducing the need for manual updates. However, both studies provided limited discussion on building maintenance. Some research has focused on developing digital twins specifically for building assets and systems. Lu~\textit{et al.}~\cite{lu_digital_2020} presented a digital twin-based anomaly detection process for assets in a building. Hosamo~\textit{et al.}~\cite{hosamo_digital_2022} focused on creating a digital twin of HVAC systems to optimize energy usage and maintain thermal comfort. They leveraged BIM for modeling complex systems and integrated IoT sensor data for real-time monitoring and performance optimization~\cite{hosamo_digital_2022}.

In the context of preserving historic buildings, some studies have also made contributions. Angjeliu~\textit{et al.}~\cite{angjeliu_development_2020} created a digital twin of a historic masonry building by integrating data such as geometry, visual observation, construction process, and material properties. The model was calibrated with dynamic measurements to assess structural integrity and support preventive maintenance. Ni~\textit{et al.}~\cite{ni_enabling_2022} proposed a digital twin solution for historic buildings, combining IoT for real-time data collection and ontology for consistent data representation. However, this solution relies on a specific cloud service provided by Microsoft Azure~\cite{azure_products_online}, i.e., Azure Digital Twins, which has no alternatives provided by other public cloud vendors. Zhang~\textit{et al.}~\cite{zhang_automatic_2022} developed a digital twin platform for optimizing RH in underground heritage sites. The platform includes modules for sensor position schematics, environmental data visualization, fan control, BIM representation, and equipment information. 

\section{Methodology}
\label{sec:methodology}

This section outlines the method for creating parametric digital twins. It begins by introducing the high-level architecture of the digital twin system, including its main components and their interactions. Next, it presents a reference implementation that includes the edge platform and IoT sensors, data modeling for contextual and time series data, data applications built on open-source software libraries, and the utilization of public cloud services.

\subsection{System Design}
\begin{figure}[!tb]
\includegraphics[width=\columnwidth]{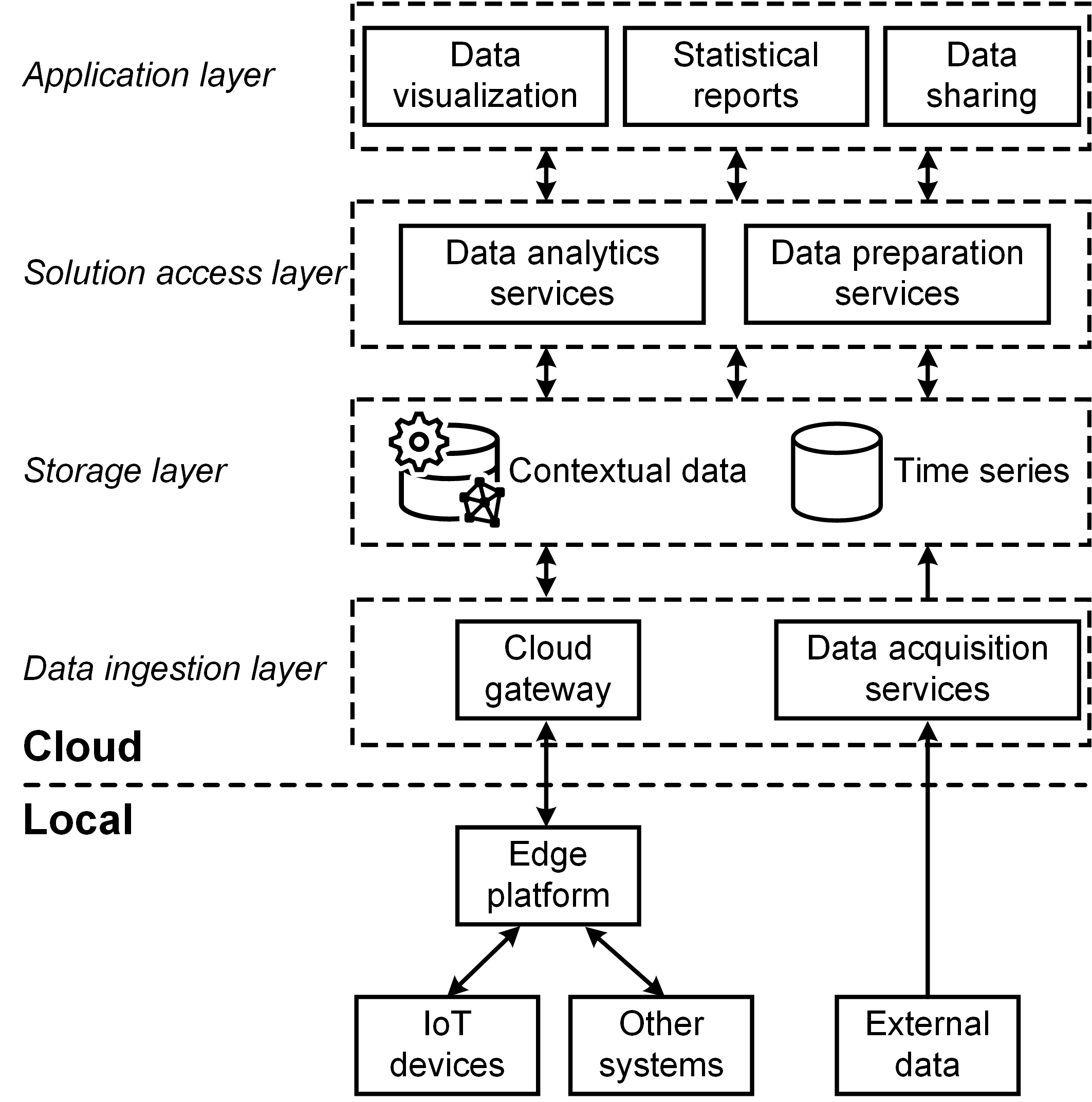}
\centering
\caption{Architecture of the digital twin system.}\label{fig:ni1}
\end{figure}

The system is designed to facilitate conservation of historic buildings by providing a comprehensive environment for data exploration. To achieve this goal, the system offers functions such as data collection, modeling, storage, querying, analysis, and visualization. As shown in Fig.~\ref{fig:ni1}, the architecture of the system comprises two parts: a local part and a cloud part.

\subsubsection{The Local Part}

The local part serves as the primary data source for the entire system, collecting both internal and external data related to building operations and maintenance. Internal data can include indoor climate, energy usage, equipment operations, system control settings, and occupant information. These data are collected using IoT devices like environmental sensors and actuators. The edge platform preprocesses these data as needed, organizes them into a predefined format, and then uploads them to the cloud. With its computing, storage, and network capabilities, the edge platform can also handle tasks previously executed in the cloud, which makes it suitable for scenarios requiring low latency or enhanced privacy. Moreover, existing systems in a building, e.g., building management system, can be integrated into the edge platform to enable the use of data from various subsystems. External data relevant to building maintenance, such as weather conditions and energy prices, are collected through open application programming interfaces (APIs). These rich data ensure that the digital twin system can effectively monitor, manage, and optimize building operations and maintenance.

\subsubsection{The Cloud Part}

The cloud part is organized into four layers from bottom to top: data ingestion layer, storage layer, solution access layer, and application layer.

Data ingestion layer handles data entry into databases. It includes a cloud gateway, which translates messages between different protocols, and data acquisition services that collect external data.

Storage layer primarily stores two types of data: contextual data and time series data. Contextual data include information about objects and their relationships, such as sensor positions and spatial relationships between rooms and floors. These data are relatively stable and are stored using a graphical data model. Time series data consist of sampled values over time and are stored in a relational database or a database optimized for time series. These two types of data are correlated using identifiers like universally unique identifiers (UUIDs).

Solution access layer manages data access. It also performs data analytics and preparation, such as calculating statistics and resampling. Services in this layer interact with the data by reading or writing historical or current data points. Due to the different formats of contextual and time series data, interactions often require two steps: searching for identifiers in contextual data and then accessing corresponding time series.

Application layer delivers user-facing applications that support decision-making for building operations and maintenance. It includes tools for visualizing collected data, generating insights through analysis, and sharing data among experts from diverse fields.

\subsection{Implementation}

The proposed implementation offers a systematic approach for creating parametric digital twins of historic buildings. It integrates multiple environmental sensors for collecting data about indoor climate, utilizes an edge platform for communication, combines graphical and relational models for organizing data, and leverages Microsoft Azure cloud services for data storage and deployment of developed applications. The implementation also considers scalability, which allows for future expansion and integration of additional features.

\subsubsection{Edge Platform and Sensors}
\label{sec:edge_sensors}

The edge platform was built using a Raspberry Pi 3 Model B+ computer (Raspberry Pi Foundation, Cambridge, GBR) paired with a Grove Base Hat (Seeed Technology, Shenzhen, CHN). This setup accommodates various communication interfaces, such as UART, I2C, as well as digital and analog IOs, enabling easy integration with a wide range of sensors. To ensure Internet connectivity in areas lacking Wi-Fi coverage, a 4G USB modem E3372-325 (Huawei, Shenzhen, CHN) was integrated with the Raspberry Pi computer.

The following sensors were used to measure six environmental parameters: 
\begin{itemize}
  \item AHT20 (Adafruit, New York, USA) measures both temperature and RH.
  \item MH-Z16 (Winsen Electronics Technology, Zhengzhou, CHN) measures CO\textsubscript{2} concentration.
  \item PPD42NS (Shinyei Corporation, New York, USA) detects particulate matter with a diameter larger than 1~\textmu{m}.
  \item LM358 (Seeed Technology, Shenzhen, CHN) senses the relative intensity of ambient noise.
  \item TSL2591 (Adafruit, New York, USA) measures light intensity.
\end{itemize}

\begin{figure}[!tb]
\includegraphics[width=\columnwidth]{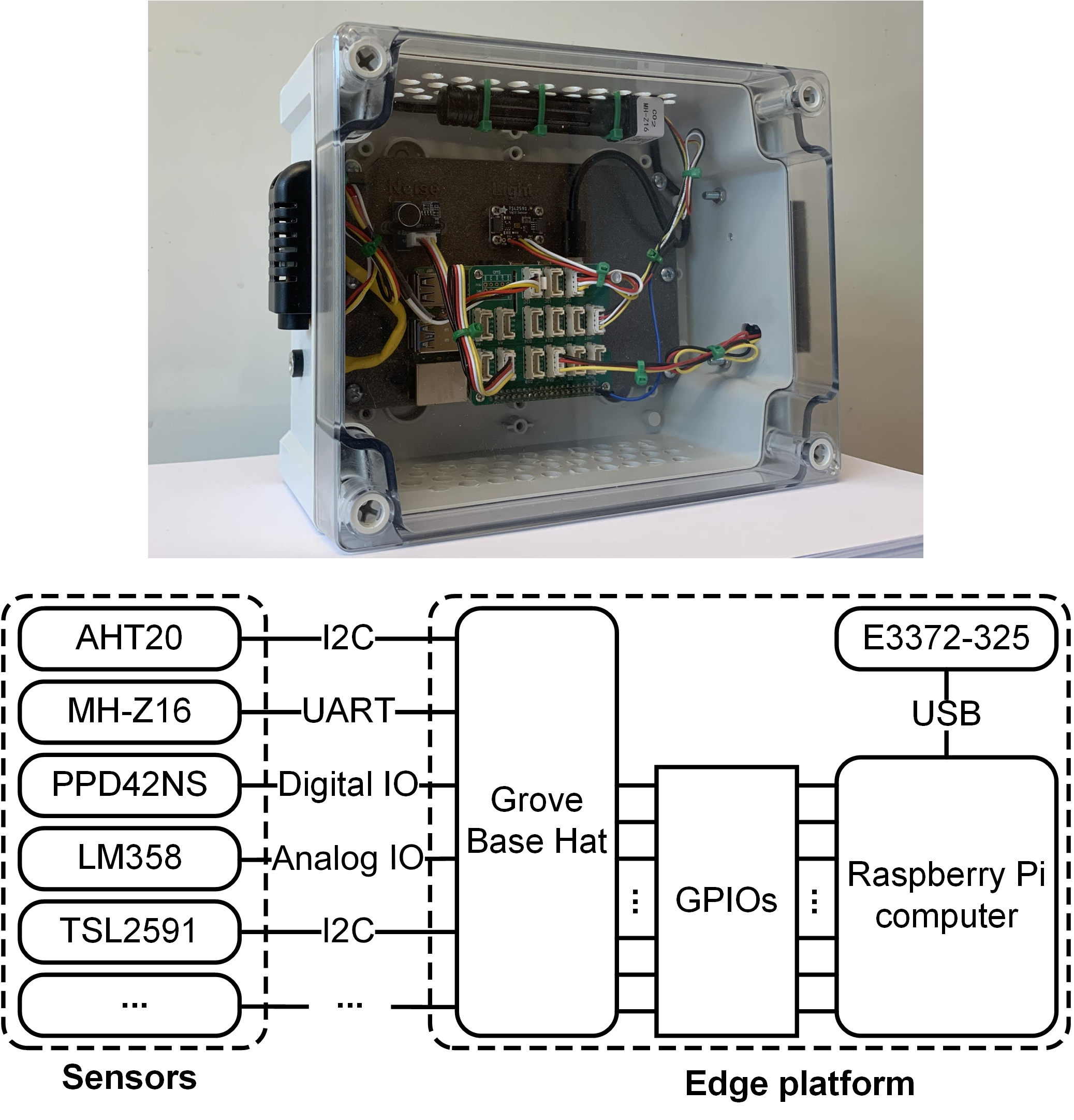}
\centering
\caption{The upper photo shows a sensor box that packages an edge platform and six sensors. The lower figure depicts hardware block diagram of used components and their connections.}\label{fig:ni2}
\end{figure}

As shown in Fig.~\ref{fig:ni2}, the edge platform and six sensors are packaged in a plastic box (size: 201~mm$\times$163~mm$\times$98~mm). More sensors can be added based on specific need. For example, in the case study described in Section~\ref{sec:case_study}, two additional ultrasonic sensors (UR18.DA0.2-UAMJ.9SF, Baumer, Frauenfeld, CHE) were added to the sensor box to monitor groundwater level changes in the basement.

The performance of these sensors in accuracy and reliability is enough for this work, but the limitations should be considered. The AHT20 sensor has reduced accuracy ($\pm$3\%) for measuring RH when the environment has an RH outside the typical range (20--80\%). The MH-Z16 sensor has declined accuracy at higher CO\textsubscript{2} concentrations ($\pm$(50 ppm $+$ 5\% of the reading)). The PPD42NS sensor detects particulate matter (PM) larger than 1~\textmu{m} but does not distinguish them, making it unsuitable for PM2.5 or PM10 detection. Lastly, the LM358 sensor provides relative noise levels without precise decibel measurements.

\subsubsection{Data Model}

Contextual data were organized by extending the Brick ontology (v1.3.0)~\cite{balaji_brick_2018}. This ontology provides semantic descriptions of physical, logical, and virtual assets within buildings, along with their interrelationships. Extending existing ontology minimizes redundancy and enhances the adaptability of our implementation to other buildings. The organized contextual data were accessed and queried using the SPARQL query language, which ensures efficient retrieval and utilization of structured building information.

Time series data, such as sensor readings, were stored in a single table within a relational database. The table consists of three columns: TIME, UUID, and VALUE. The primary key is a composite of TIME and UUID. UUID uniquely identifies a time series, while TIME and VALUE are the timestamp and value of a measurement, respectively. This design ensures efficient storage and retrieval of temporal data associated with unique identifiers, which facilitates robust data management and analysis.

\subsubsection{Data Applications}

Data applications, including data visualization, statistical reporting, and data sharing, were developed using Python (v3.8.18) and libraries: pandas (v1.4.1), Dash (v2.3.1), and Dash Bootstrap Components (v0.13.1). Pandas is widely recognized for its efficiency, flexibility, and ease in handling data analysis and manipulation tasks. Dash is an open-source and low-code framework. It enables rapid development and deployment of data applications with customizable user interfaces, simplifying the creation of full-stack web apps with interactive data visualizations. Dash Bootstrap Components enhances Dash by providing Bootstrap components specifically designed for Plotly Dash, allowing for responsive layouts with consistent styling in complex applications.

\subsubsection{Public Cloud Services}

Several public cloud services provided by Microsoft Azure~\cite{azure_products_online} were utilized to implement the cloud components. These services include Azure IoT Hub, Azure Functions, Azure SQL Database, and App Service. Azure IoT Hub acts as the cloud gateway to enable communication between the cloud and connected edge platforms. Azure Functions provides event-driven serverless computing capabilities. It was used to implement functions like writing data to the database and retrieving data from external sources. Azure SQL Database is a cloud-based relational database service known for its automatic scalability and high availability. It was utilized to store time series data. App Service was used to host developed data applications. It is important to note that the cloud services mentioned above have alternatives from other providers, e.g., Amazon Web Services~\cite{aws_iot_twin_maker_online}, which helps avoid deep binding with a single cloud vendor.
 
\section{Case Study}
\label{sec:case_study}

This section presents an application of the implemented system in a historic building. First, a brief description of the building is provided. Then, deployment of local devices and collection of external data are illustrated. After that, created data applications are introduced. Finally, it describes methods used to evaluate indoor climate in the building.

\subsection{Description of Löfstad Castle}

\begin{figure}[!tb]
\includegraphics[width=\columnwidth]{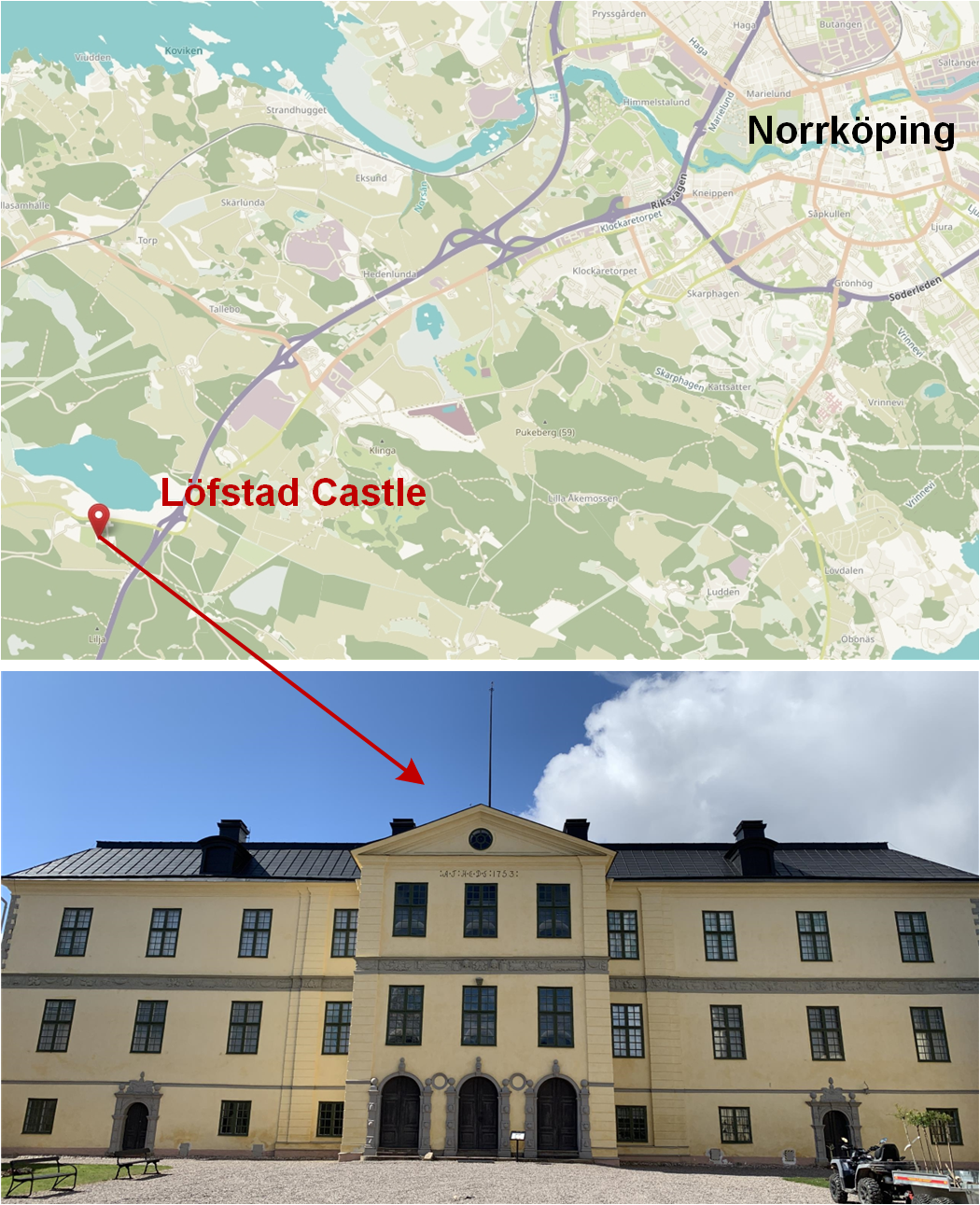}
\centering
\caption{The upper map depicts the location of Löfstad Castle, while the lower photo shows the main building of the castle.}\label{fig:ni3}
\end{figure}

Löfstad Castle is located about 9~km southwest of the city of Norrköping in Östergötland, Sweden. The main building of the castle (see Fig.~\ref{fig:ni3}) has a basement, three floors above the basement (ground, first, and second floors), and an attic. The entrance of the main building faces east, towards a courtyard flanked by two wings on the north and south sides. The main building was constructed with heavy masonry, with outer walls approximately 1.1~m thick. Currently, the main building is naturally ventilated, and only the first floor has electric radiant heating during the winter season.

Löfstad Castle holds significant cultural and historic values. It has been a designated historic monument since 1983~\cite{lofstad_castle_swedish_national_heritage_board_online}. The castle was first built in the 17th century. In 1750, a devastating fire broke out. Both the main building and the north wing were severely damaged while the south wing was largely unaffected. The castle was then rebuilt and got its current form with a Baroque characteristic and some Rococo elements~\cite{hedlund_2013}. Miss Emilie Piper (1857--1926) was the last owner of Löfstad Castle~\cite{lofstad_kulturmiljoenheten_online}. In her will (1923), she donated the majority of the collections in the castle to Östergötland's Museum, and the estate property to Riddarhuset. She also stated that the interior of the stateroom on the first floor should be preserved exactly as it was. Today, the estate property is managed by Riddarhuset through a foundation. The Löfstad Castle Museum has been open to the public since 1942. Over the years, only minimal changes have occurred to the interior of the stateroom and the collection it houses. Other parts of the interior of the main building have slightly changed, but its historic character has been mostly retained~\cite{en_bok_om_lofstad_slott_2022}. These efforts make the castle a well-preserved example of manor buildings from the 17th and 18th centuries~\cite{lofstad_kulturmiljoenheten_online}.

\begin{figure}[!tb]
\includegraphics[width=\columnwidth]{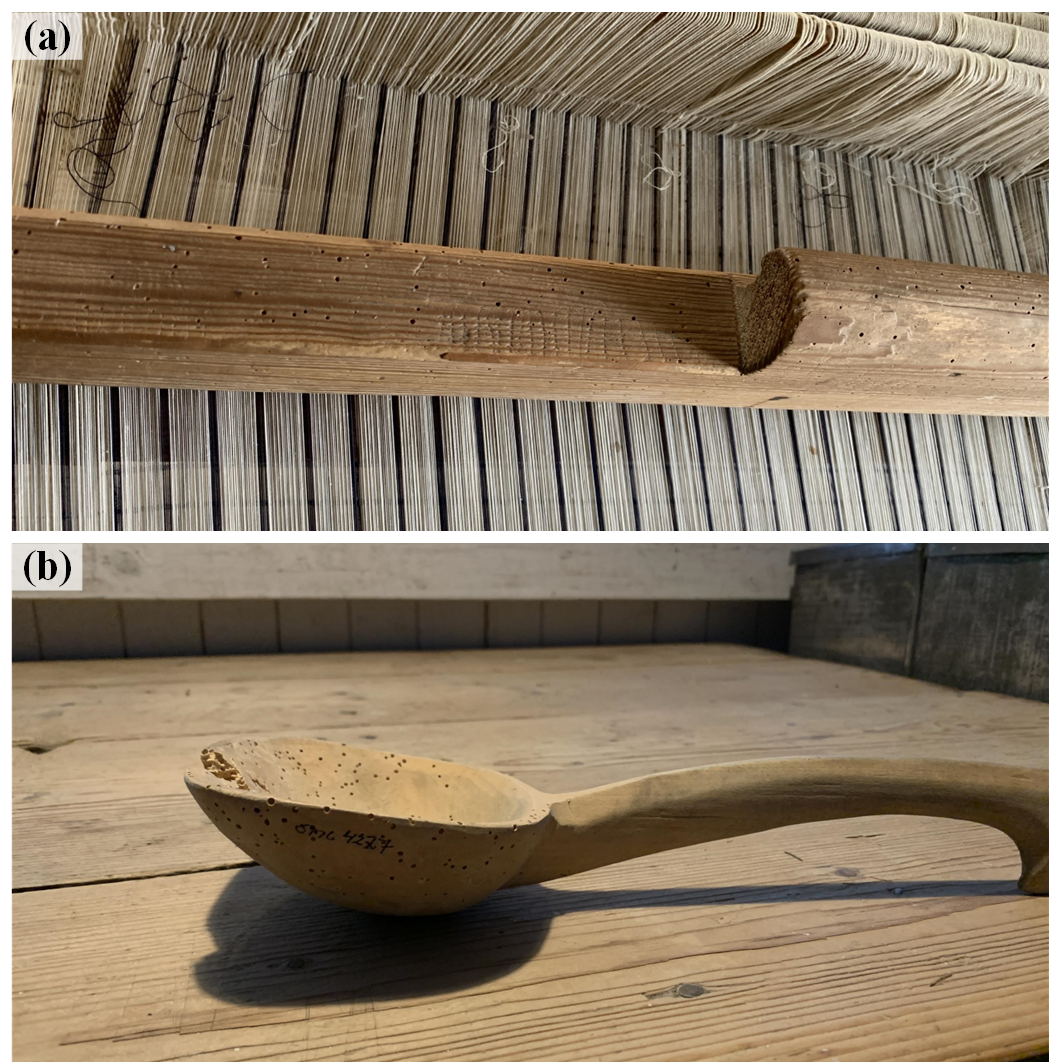}
\centering
\caption{Examples of wooden objects damaged by woodworms: (\textbf{a}) beam of a loom and (\textbf{b}) a wooden ladle.}\label{fig:ni4}
\end{figure}

However, the historic interior of the main building and housed collections are deteriorating due to inappropriate indoor climate, primarily high RH. The problem originates in the basement, where the floor consists of exposed soil, allowing groundwater to evaporate from the soil into the air continuously. This moisture is speculated to rise through walls, increasing humidity levels on upper floors. High RH poses risks of mold and other biological damage. The high humidity creates ideal conditions for woodworm, which causes considerable harm to wooden objects in the building (see Fig.~\ref{fig:ni4}).

\subsection{Deployment of Local Devices}

\begin{figure*}[!tb]
\includegraphics[width=\textwidth]{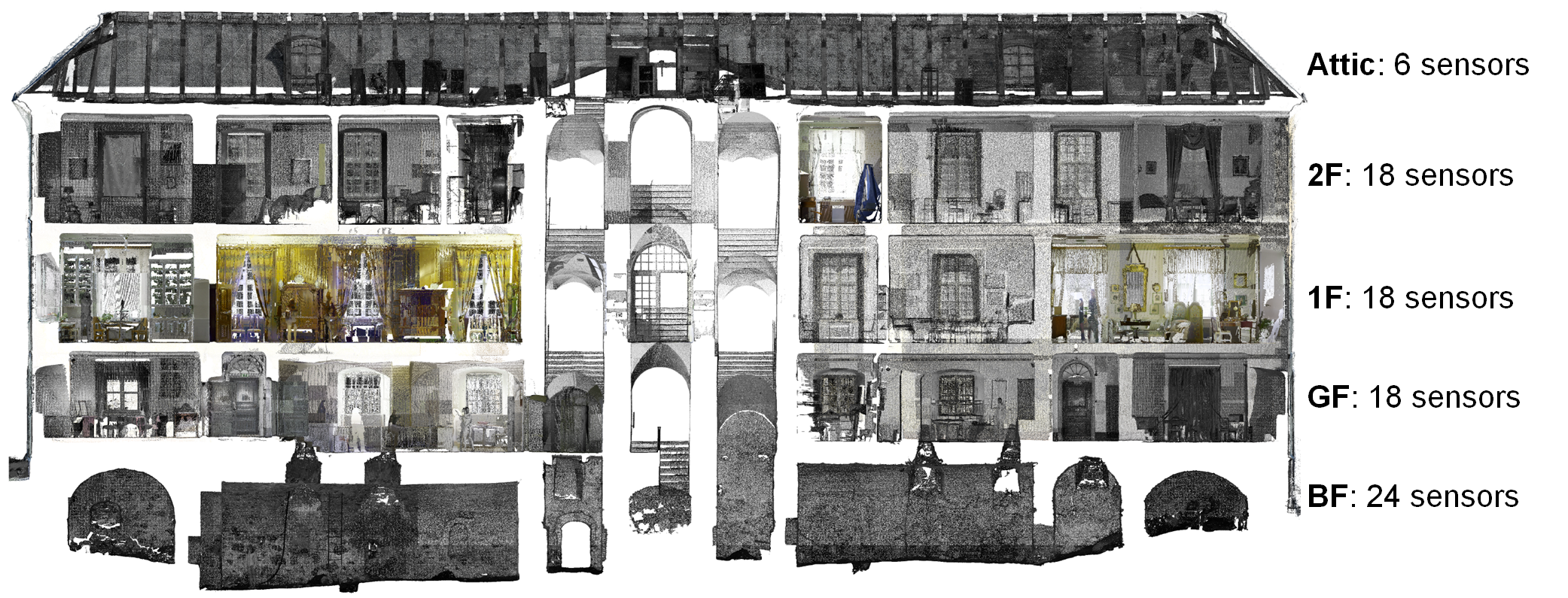}
\centering
\caption{A section drawing of the 3D model of the main building and an overview of sensor deployment on each floor. The basement, ground floor, first floor, and second floor are abbreviated as BF, GF, 1F, and 2F, respectively.}\label{fig:ni5}
\end{figure*}

As shown in Fig.~\ref{fig:ni5}, a total of 84 sensors are distributed throughout the building, from the basement to the attic, to continuously monitor indoor environmental conditions on all floors. The number of sensor boxes deployed in the basement (BF), on the ground floor (GF), first floor (1F), second floor (2F), and attic is three, three, three, three, and one, respectively, with each sensor box containing sensors as described in Section~\ref{sec:edge_sensors}.

\begin{figure*}[!tb]
\includegraphics[width=\textwidth]{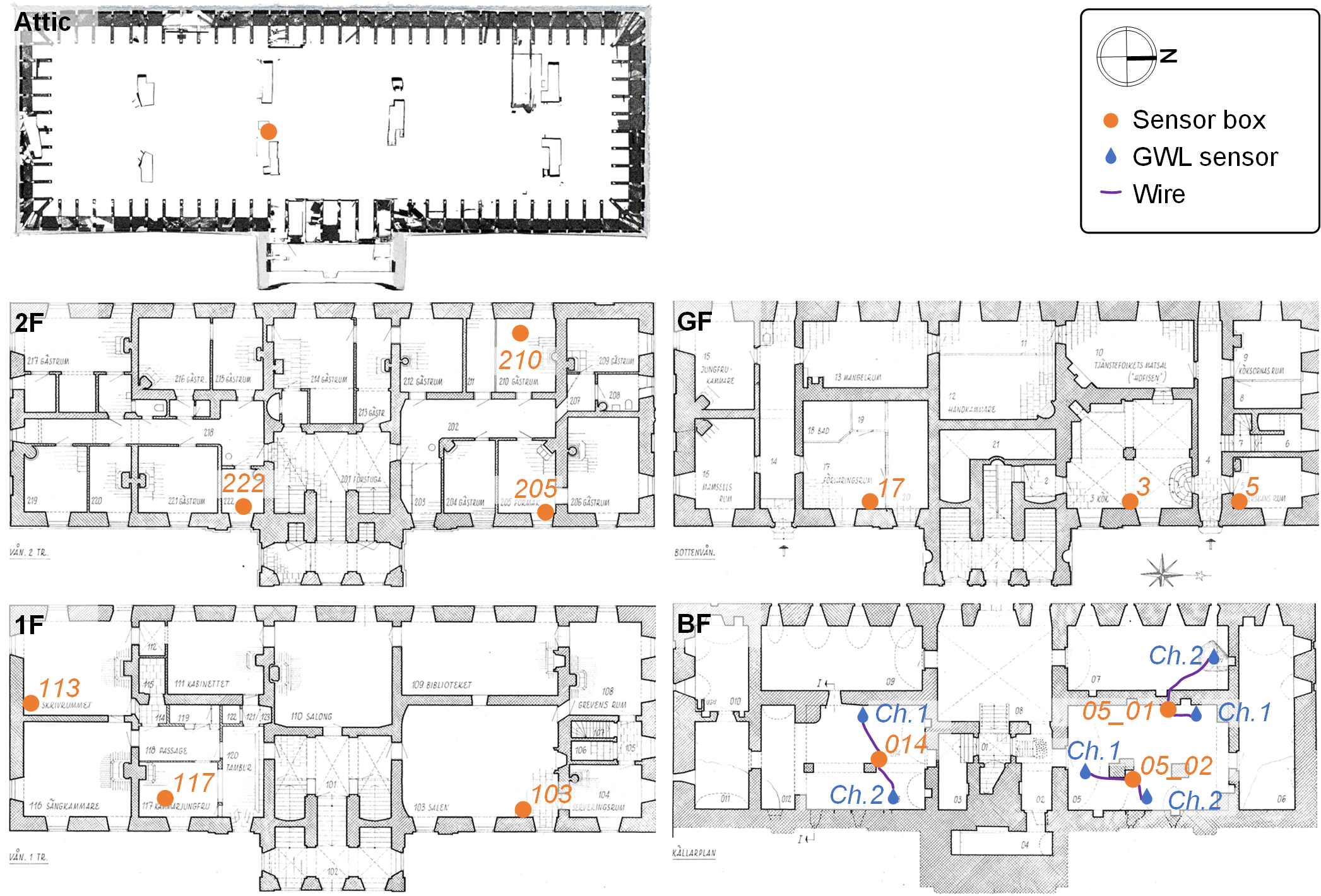}
\centering
\caption{Placements of sensor boxes and groundwater level (GWL) sensors. Orange dots represent sensor box locations, while blue drops mark GWL sensor positions. Numbers next to orange dots are room identities. In the basement, each sensor box has two additional GWL sensors connected, labeled as Channel 1 (Ch. 1) and Channel 2 (Ch. 2).}\label{fig:ni6}
\end{figure*}

Placements of sensor boxes and groundwater level (GWL) sensors are indicated on floor plans of the main building (see Fig.~\ref{fig:ni6}). Details on the location, height above the floor, and installation date of sensor boxes are provided in Table~\ref{tab:table1}. All six GWL sensors were also installed on June 28, 2023. The positioning of sensor boxes was carefully chosen to meet both building conservation requirements and site conditions. The considered criteria include:
\begin{itemize}
    \item Strategic coverage: Sensor boxes were primarily distributed in rooms containing valuable collections, with a secondary emphasis on ensuring even coverage across rooms on different floors. This helps to understand how environmental conditions vary throughout the building. 
    \item Minimizing interference: Sensor boxes were deliberately positioned away from heat sources, such as radiators, to avoid measurement bias.
    \item Monitoring vulnerable areas: Special attention was given to locations susceptible to environmental fluctuations, including areas near windows, exterior walls, and doors, to effectively capture changes in temperature and humidity.
    \item Height considerations: In the absence of other considerations, sensor boxes were placed at approximately eye level (1.5--2 m) to avoid skewed measurements.
    \item Non-intrusive installation: To preserve structural and historical integrity of the building, sensor boxes were installed without causing damage to walls and floors.
\end{itemize}

The data continuously collected by these sensors are essential for creating a parametric digital twin and evaluating indoor climate of the building. Specifically, the sensors capture dynamic environmental parameters, which are then integrated into the digital twin to create a detailed and constantly updated model of the building. This enables real-time monitoring and provides a data-driven approach to support proactive decision-making for historic building conservation.

\begin{figure}[!tb]
\includegraphics[width=\columnwidth]{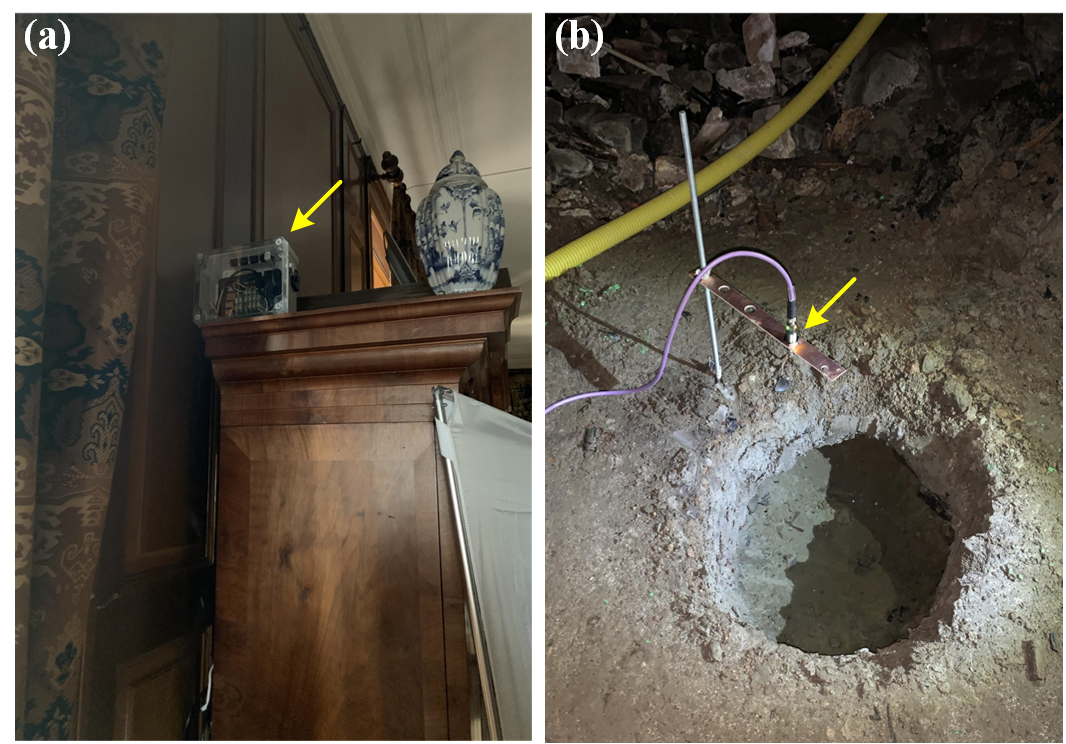}
\centering
\caption{Examples of deployment: (\textbf{a}) sensor box deployed in Room 103 on the first floor, and (\textbf{b}) GWL sensor (Ch. 1) connected to sensor box 05\_01 in the basement.}\label{fig:ni7}
\end{figure}

Examples of a deployed sensor box and a GWL sensor are illustrated in Fig.~\ref{fig:ni7}. In Room 103 (the stateroom) on the first floor, the sensor box is placed on top of a cabinet (see Fig.~\ref{fig:ni7}{a}). This sensor box provides insights into both indoor climate and occupancy pattern of the room. Fig.~\ref{fig:ni7}{b} shows a GWL sensor, which uses ultrasound to measure the distance from the water surface to the probe. These measurements of GWL are integrated into the parametric digital twin, where they are analyzed alongside outdoor precipitation to identify correlations between rainfall and groundwater levels. Understanding this relationship helps to evaluate potential risks such as rising dampness. This insight facilitates implementing proactive conservation strategies within the digital twin framework, such as optimizing humidity control, to preserve the building.

\begin{table}[!tb]
\centering
\caption{The location, height above the floor, and installation date of deployed sensor boxes. A height of zero means the sensor box is put directly on the floor.}
\label{tab:table1}
\begin{tabular}{@{}rlrr@{}}
\toprule
\multicolumn{2}{c}{Location} & \multicolumn{1}{c}{Height (m)} & \multicolumn{1}{c}{Date} \\ \midrule
Attic &  & 2.9 & 20/04/2023 \\
 &  &  &  \\
\multirow{3}{*}{2F} & Room 205 & 0.9 & 13/01/2023 \\
 & Room 210 & 1 & 14/02/2023 \\
 & Room 222 & 0.8 & 14/02/2023 \\
 &  &  &  \\
\multirow{3}{*}{1F} & Room 103 & 2 & 13/01/2023 \\
 & Room 113 & 0.4 & 20/04/2023 \\
 & Room 117 & 0.8 & 13/01/2023 \\
 &  &  &  \\
\multirow{3}{*}{GF} & Room 3 & 0.9 & 13/01/2023 \\
 & Room 5 & 0 & 29/06/2022 \\
 & Room 17 & 2.4 & 14/02/2023 \\
 &  &  &  \\
\multirow{3}{*}{BF} & Room 05\_01 & 2.1 & 14/02/2023 \\
 & Room 05\_02 & 0 & 28/06/2023 \\
 & Room 014 & 0 & 13/01/2023 \\ \bottomrule
\end{tabular}
\end{table}

All sensors were calibrated before deployment. Sensors for measuring temperature, RH, and GWL were calibrated by ourselves. The remaining sensors were calibrated by manufacturers. Each environmental parameter is collected every 30 seconds (two samples per minute) to capture rapid fluctuations, with the possibility for downsampling during subsequent data analytics and visualization.

\subsection{Collection of External Data}

Collected external data are outdoor weather conditions that affect indoor climate, including dry-bulb temperature, RH, dew point temperature, precipitation, and air pressure. Hourly measurements of these parameters were obtained through open APIs provided by the Swedish Meteorological and Hydrological Institute~\cite{smhi_open_data_online}. The data were measured from a weather station (ID: 86340) located about 7~km from Löfstad Castle. These measurements serve as an approximation of ambient weather conditions surrounding the building. The data obtained from the weather station were considered sufficient for this work since the deviations were negligible. While deploying environmental sensors outside the building would provide more localized and accurate ambient weather data, it should fulfill constraints for historic building conservation to minimize any external alterations that could affect its aesthetics or structure.

\subsection{Developed Applications}

\begin{figure*}[!tb]
\centering
\subfloat[]{\includegraphics[width=0.8\textwidth]{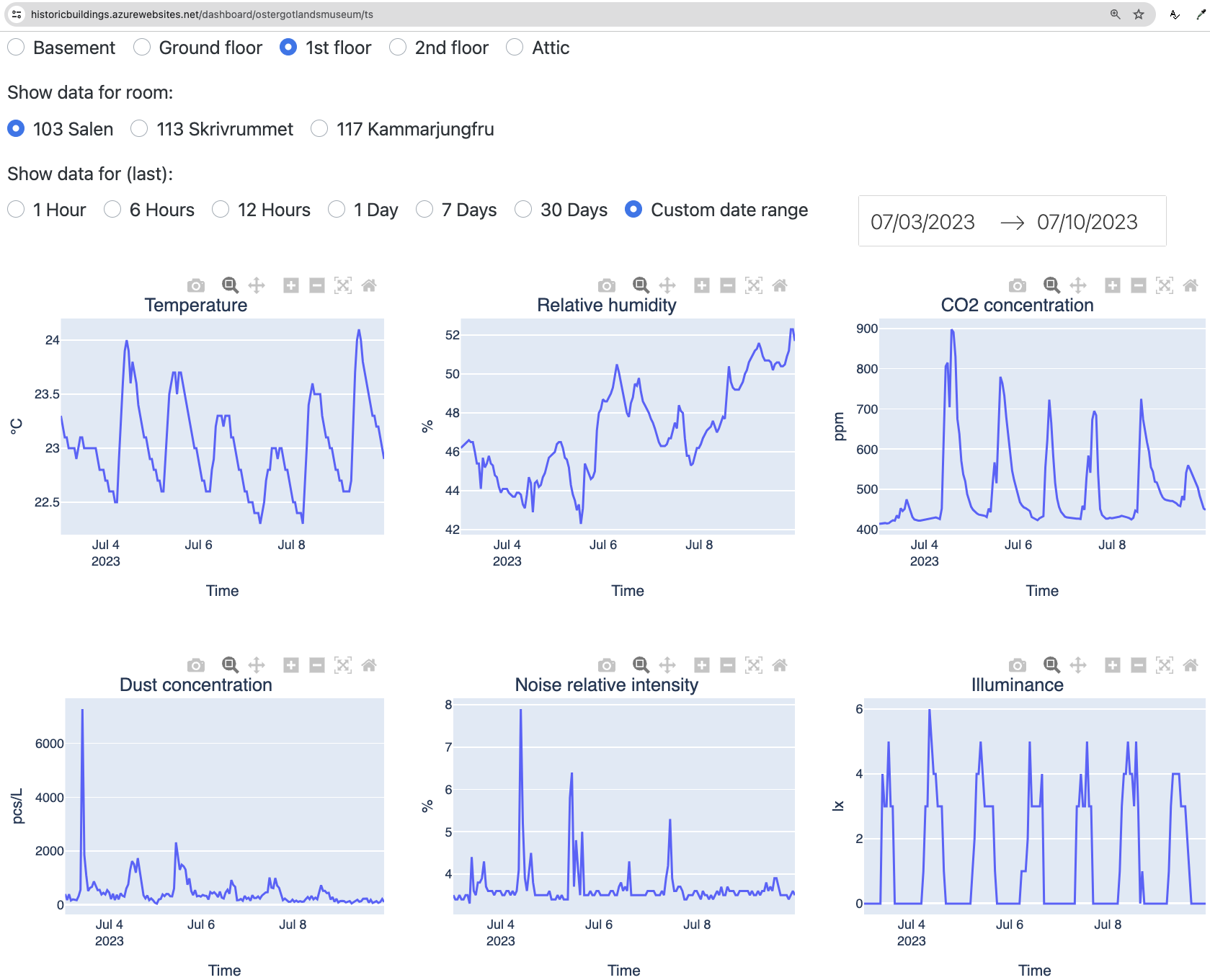}}
\hfill
\subfloat[]{\includegraphics[width=0.8\textwidth]{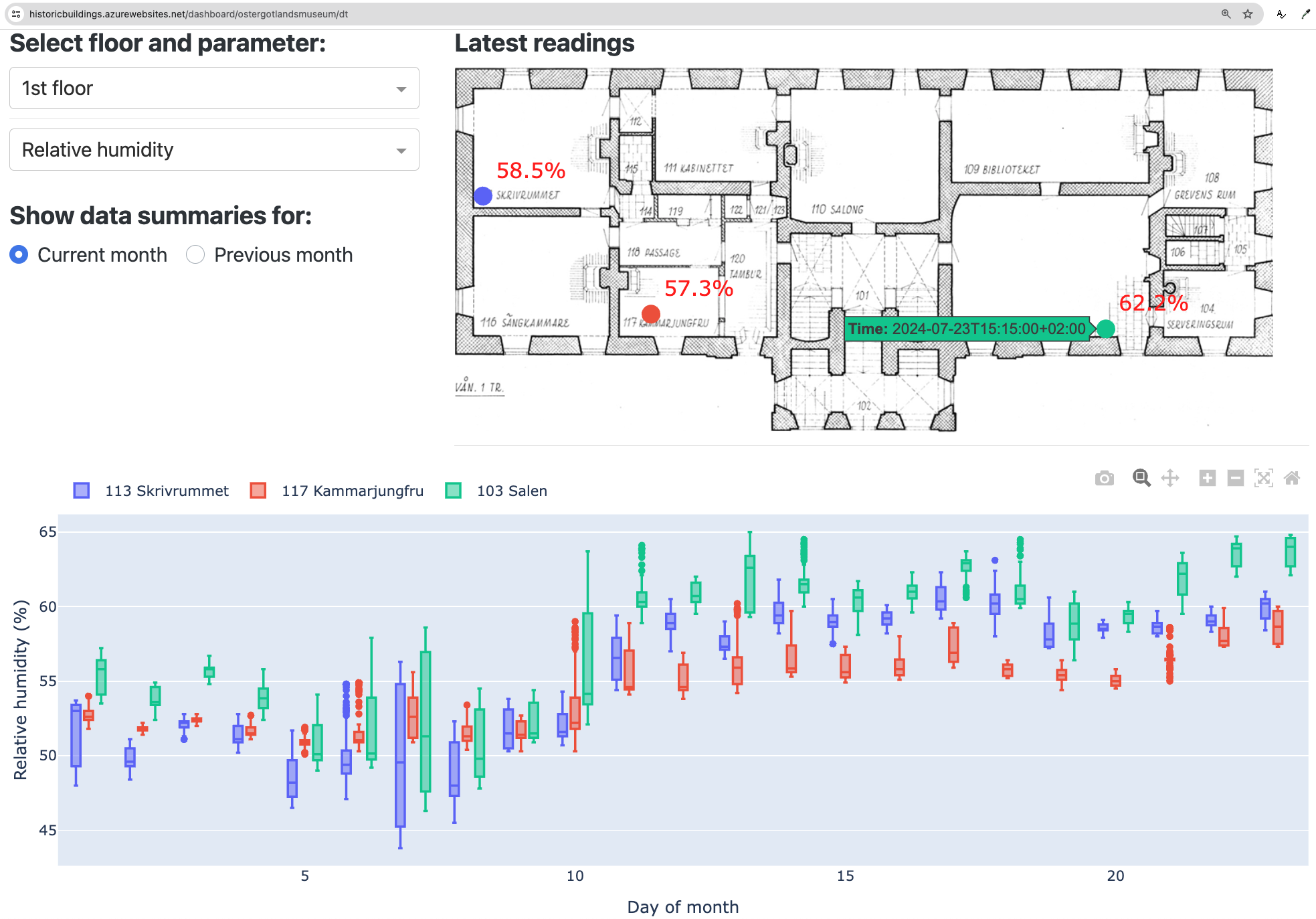}}
\caption{(\textbf{a}) The data application to display time plots of collected indoor environmental data based on selected room and time period. (\textbf{b}) Data applications to show the latest measurements of a selected parameter across a chosen floor, along with daily distributions for a selected month.}
\label{fig:ni8}
\end{figure*}

Four data applications were developed, focusing on data visualization, analytics, and sharing. Fig.~\ref{fig:ni8}{a} presents the application to visualize time plots of environmental parameters collected from a selected sensor box within a specified date range. This tool helps users inspect individual monitoring points, identify trends and fluctuations, as well as explore correlations between different environmental parameters. For example, based on the one-week data shown in Fig.~\ref{fig:ni8}{a}, it is evident that temperature exhibited strong daily seasonality, while CO\textsubscript{2} concentration showed six distinct peaks, indicating different levels of occupancy in Room 103 on those days. Moreover, during each day of the five days from July 4 to 8, a noticeable drop followed by an increase in temperature was observed around midday (see Fig.~\ref{fig:ni8}{a}). This fluctuation is speculated to be associated with human activity, especially a temporary leave followed by a return. Pearson correlation coefficient between temperature and CO\textsubscript{2} concentration over the five days is 0.64, indicating a strong positive correlation~\cite{evans_straightforward_1996} between these two parameters, which suggests that the observed temperature variations are correlated with occupancy patterns.

In addition to time plots, two data applications, as shown in Fig.~\ref{fig:ni8}{b}, provide a comprehensive view of collected data. The upper part displays the latest measurements of a selected environmental parameter directly on the floor plan, enabling easy comparison across different locations on the same floor. The lower part of Fig.~\ref{fig:ni8}{b} uses a boxplot to illustrate the daily distribution of a selected environmental parameter for a chosen month (current or previous). This visualization aids in understanding daily fluctuations and detecting abnormal changes in environmental parameters over time.

Open data access allows project participants and external researchers to obtain data for further analysis from various perspectives. The data sharing application (available at https://historicbuildings.azurewebsites.net/download) enables users to choose various sampling intervals (raw or downsampled), parameters, and time periods. Downloaded data are cleaned of two types of anomalies. The first type is outliers, characterized by unexpected spikes or drops in the data, typically resulting from temporary sensor malfunctions or external interference. To address these outliers, threshold-based filtering was applied to remove data points that exceed predefined upper and lower bounds. The second type involves missing data, primarily due to communication issues. To fill in these missing data, linear interpolation was employed to estimate the missing values based on adjacent data points. Cleaned data are provided in comma-separated values (CSV) format, making it accessible for analysis with various tools.

Compared to existing digital twin platforms, these data applications offer greater customization and flexibility, allowing users to visualize specific environmental parameters and analyze data from selected sensor boxes within customizable date ranges. They provide real-time insights for quick identification of trends and correlations. Additionally, user-friendly interfaces simplify complex datasets, making them accessible to stakeholders with varying technical expertise, thus supporting effective management and decision-making for historic building conservation.

\subsection{Analysis of Indoor Climate}

Due to problems with indoor climate in the building, particularly high RH in the basement and on the ground floor, a detailed analysis was conducted to better understand these conditions. The analysis aims to answer the following questions:
\begin{enumerate}
    \item What are the patterns of temperature and RH at each monitoring point?
    \item How does the high RH in the basement influence the upper floors?
    \item How do GWL changes vary across different locations in the basement?
    \item What impact does presence of occupants have on various indoor environmental parameters?
\end{enumerate}

To address question 1, indoor and outdoor temperature and RH throughout a year were explored to identify trends, levels, and fluctuations. Temperature and RH across different rooms on the same floor and between rooms on different floors were also compared to investigate similarities and differences.

For question 2, indoor and outdoor humidity mixing ratios (MRs) over a year were compared to identify potential moisture sources. The MR is a measure of how much moisture is present in the air relative to the amount of dry air, expressed in grams of water vapor per kilogram of dry air (g/kg)~\cite{en_16242_conservation_2012}. This ratio provides a clear understanding of moisture content in the air. A comparison between indoor and outdoor MRs suggests a water source or sink in a building~\cite{brostrom_climate_2015}. If the average indoor MR consistently exceeds the outdoor MR, it might indicate constant evaporation from the ground or walls of the building. MR is calculated by (\ref{eq:mr}).
\begin{equation}
\text{MR}\ =\ 38.015\times \frac{10^{\frac{7.65t}{243.12+t}} \times \text{RH}}{p-\left( 0.06112\times 10^{\frac{7.65t}{243.12+t}} \times \text{RH}\right)},\label{eq:mr}
\end{equation}
where $t$ is temperature ($^{\circ}$C), RH is relative humidity (\%), $p$ is atmospheric pressure (hPa). Since indoor atmospheric pressure was not measured, $p$ was substituted with a standard pressure of 1013 hPa for approximate calculations of indoor MRs, as recommended by the standard EN 16242:2012~\cite{en_16242_conservation_2012}. Using the standard pressure as an approximation for indoor air pressure would slightly reduce the calculated indoor MRs. However, the resulting error is below 1\%, which is within an acceptable range and does not impact comparisons between indoor and outdoor MRs. Outdoor MRs were calculated using collected atmospheric pressure.

In addition, potential consequences of inappropriate RH, such as mold growth and material deterioration caused by strain-stress cycles, were evaluated. Mold risk was assessed using the isopleth system for substrate category I (LIM\textsubscript{I})~\cite{sedlbauer_prediction_2001}. The LIM\textsubscript{I} curve was calculated using (\ref{eq:rh_lim}), as implemented in the hygrothermal simulation software WUFI-Bio\cite{lim_expression_online}.
\begin{equation}
    \text{LIM}_{\text{I}} =\ \text{cosh}( 0.128324\times ( 30-t)) +75,\label{eq:rh_lim}
\end{equation}
where $t$ is temperature ($^{\circ}$C). The calculated value is expressed as a percentage (\%). A RH value exceeding the LIM\textsubscript{I} at a given temperature is considered risky.

Short-term fluctuations of RH throughout a year were examined to determine the target range of RH and identify risks associated with strain-stress cycles, following the standard EN 15757:2010~\cite{en_15757_conservation_2010}. Three key statistics representing the average levels and variability of RH were calculated: (1) the annual average RH, (2) seasonal cycles, and (3) short-term fluctuations. The annual average is the arithmetic mean of RH readings over one year. Seasonal cycles were determined by calculating a 30-day centered moving average (CMA), which computes the mean RH for each reading based on the 15 days preceding and following the measurement. Short-term fluctuations were calculated as the deviation between the current RH reading and the 30-day CMA, accounting for both natural seasonal variations and the stress-relaxation time constant of the materials. A safe band is determined by the 7th and 93rd percentiles of recorded RH fluctuations over the monitoring period, identifying the 14\% of the most extreme and potentially risky fluctuations. If RH variations from the seasonal level are within 10\%, a deviation of up to 10\% from the seasonal RH level is considered acceptable as the safe band.

To address question 3, GWL changes at different monitoring points were compared. The impact of precipitation on GWL changes was also studied.

For question 4, the impact of human activities on indoor environmental parameters was investigated. The analysis also assessed whether natural ventilation alone was sufficient to maintain good air quality during these activities.

The Pearson correlation coefficient ($r$) was used to measure the strength of a linear relationship between paired data. The magnitude of $r$ was classified as in study~\cite{evans_straightforward_1996}:
\begin{itemize}
    \item $|r| < 0.20$: very weak
    \item $0.20 \leq |r|  < 0.40$: weak
    \item $0.40 \leq |r| < 0.60$: moderate
    \item $0.60 \leq |r|  < 0.8$: strong
    \item $|r| \geq 0.80$: very strong
\end{itemize}

The Mann-Whitney U test~\cite{mann_test_1947} was used to assess differences between two groups of data, either in distribution or median. If the distributions differed in shape, the test evaluated differences in distribution. If the distributions had the same shape, it assessed differences in the median. A significance level of 0.05 was used. Statistical analysis was conducted using Python libraries pandas (v2.2.2) and SciPy (v1.13.1).

To align with the time granularity of outdoor weather data, indoor environmental data were downsampled to an hourly frequency by averaging the values. 

\section{Results and Discussion}
\label{sec:results_and_discussion}

This section summarizes the obtained results and findings from analysis of indoor climate. It begins with an exploration of indoor temperature and RH throughout a year. Next, a comparison between indoor and outdoor MRs is presented to identify moisture sources. After that, mold risk and short-term fluctuations in RH are evaluated. Then, the impact of precipitation on GWL changes is analyzed. Finally, the impact of presence of occupants on indoor climate is discussed.

\subsection{Temperature and Relative Humidity Throughout a Year}

\begin{figure}[!tb] 
\centering
\subfloat[]{\includegraphics[width=\columnwidth]{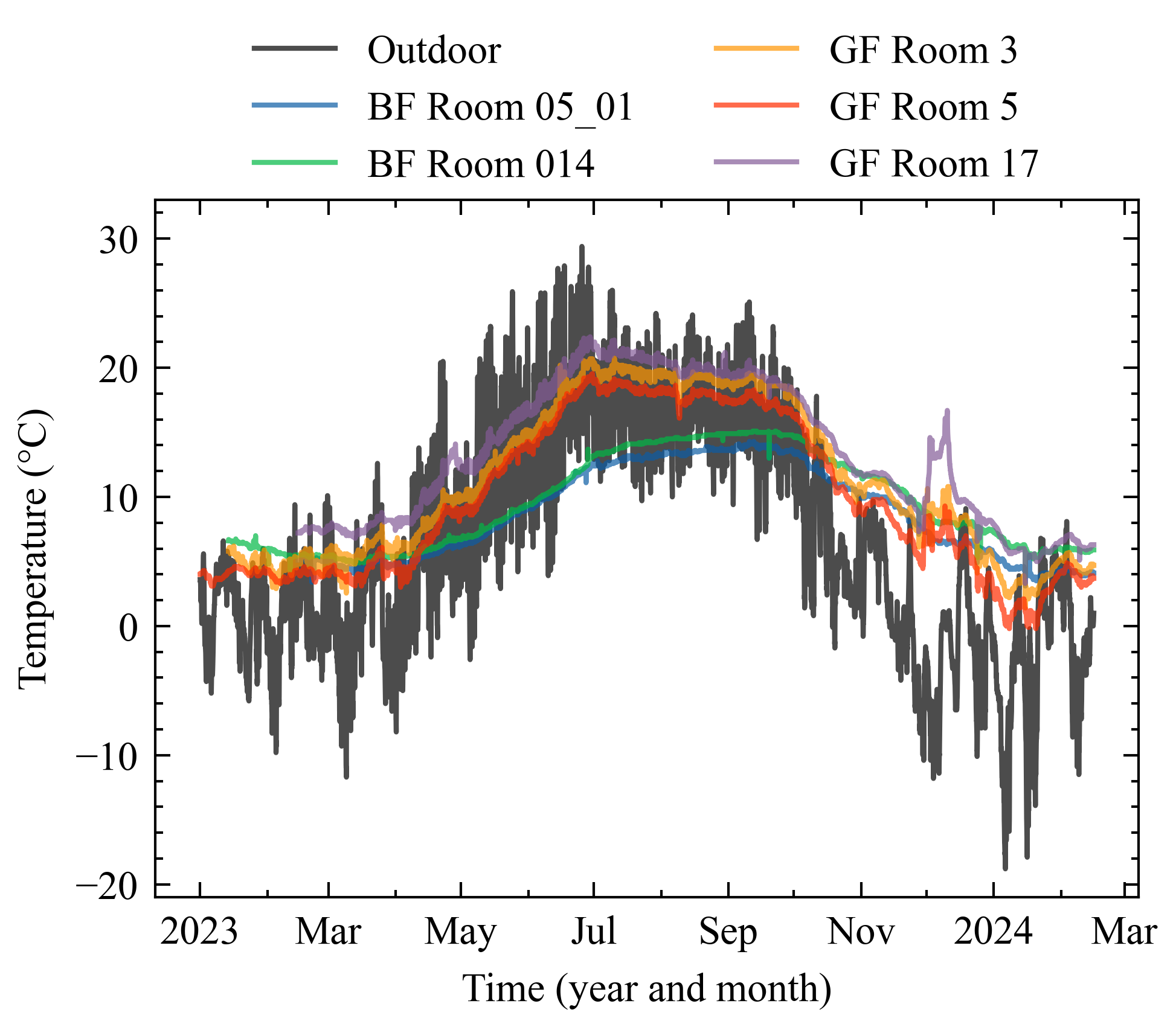}}
\hfill
\subfloat[]{\includegraphics[width=\columnwidth]{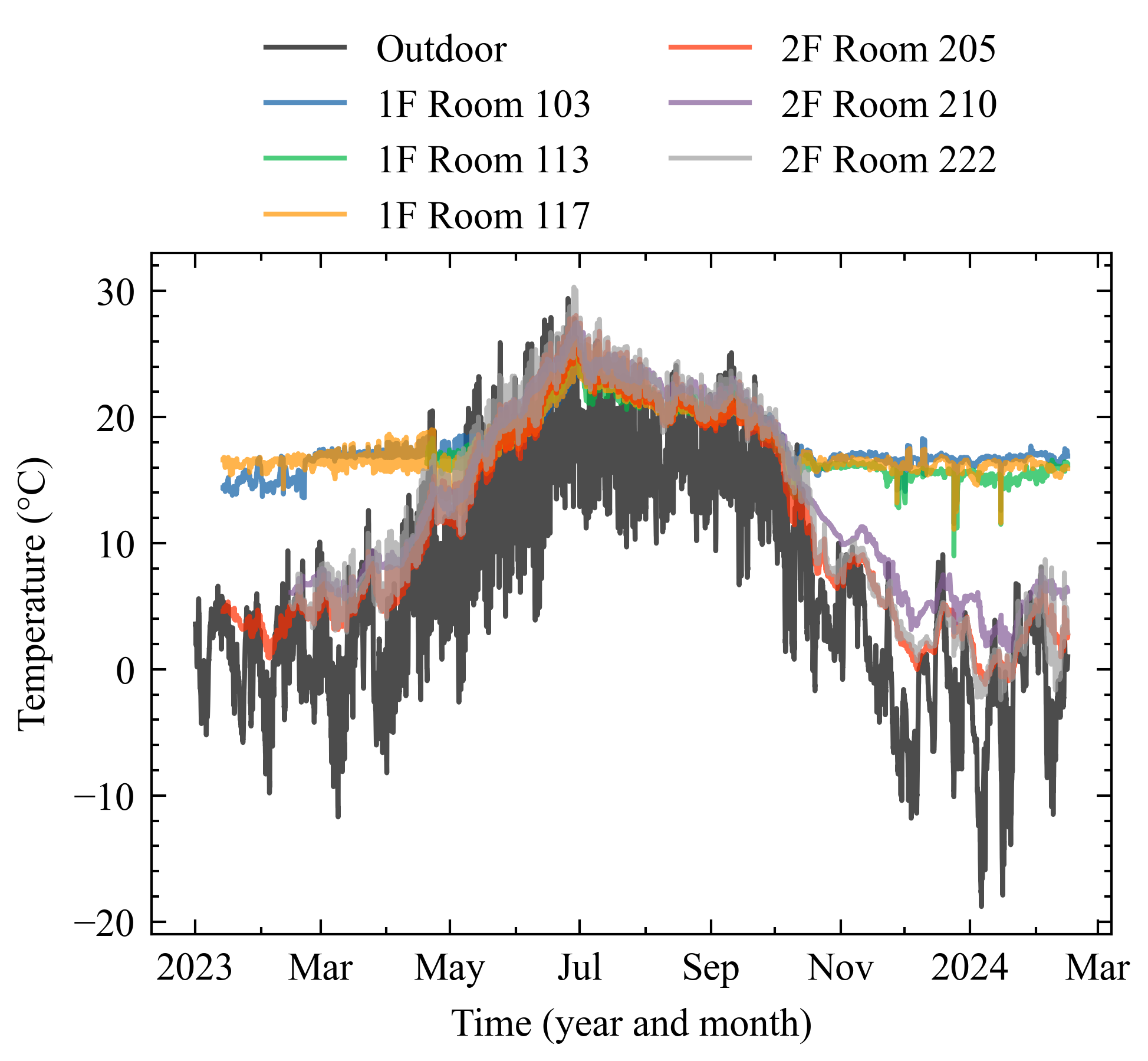}}
\caption{Historical hourly outdoor and indoor temperature of (\textbf{a}) rooms in the basement and on the ground floor as well as (\textbf{b}) rooms on the first and second floors from January 2023 to February 2024.}
\label{fig:ni9}
\end{figure}

In rooms not equipped with heating systems, the trend of indoor temperature closely follows outdoor temperature. In the main building of Löfstad Castle, rooms in the basement and rooms on the ground and second floors lack heating systems. As shown in Fig.~\ref{fig:ni9}{a} and Fig.~\ref{fig:ni9}{b}, temperature in these rooms exhibited a similar trend to outdoor temperature throughout the year. For example, there was a very strong positive correlation between indoor temperature of Room 3 and the outdoor temperature ($r = 0.84$). This suggests that the variations in indoor temperature in Room 3 were largely driven by fluctuations in outdoor temperature. Rooms on the first floor are equipped with heating systems. When these systems were turned off, the trend of indoor temperature was also driven by outdoor temperature. This pattern was evident from May to September 2023 (see Fig.~\ref{fig:ni9}{b}) for rooms on the first floor. However, when heating systems were working during the heating season, these rooms maintained a stable temperature around 16~$^{\circ}$C.

Although the trend of indoor temperature aligns with outdoor temperature, the fluctuation of indoor temperature is much smaller. As illustrated in Fig.~\ref{fig:ni9}{a} and Fig.~\ref{fig:ni9}{b}, the fluctuation of indoor temperature on all floors were considerably smoother compared to outdoor temperature. Temperature of rooms in the basement showed the least fluctuation. This reduced fluctuation for indoor temperature on all floors is owing to thick masonry outer walls of the building ($\sim$1.1~m thick), which act as a thermal buffer, moderating indoor temperature changes.

Indoor temperature of different rooms on the same floor tend to be relatively close, while noticeable differences occur between rooms on different floors. During summer months, indoor temperature generally increases as a floor goes up. As depicted in Fig.~\ref{fig:ni9}{b}, around early July 2023, temperature in rooms on the first and second floors was approximately 25~$^{\circ}$C. Temperature of Room 222 on the second floor even reached up to 30~$^{\circ}$C. In contrast, during the same period, temperature of rooms on the ground floor and in the basement were approximately 20~$^{\circ}$C and 12~$^{\circ}$C, respectively (see Fig.~\ref{fig:ni9}{a}). The Mann-Whitney U test results indicated a significant difference in indoor temperature between rooms on different floors ($p < 0.001$).

Abnormal variations in indoor temperature were observed during specific periods. As shown in Fig.~\ref{fig:ni9}{a}, Room 17 (without heating) on the ground floor experienced an unusually high temperature in the first half of December 2023. Moreover, heated rooms like Rooms 113 and 117 exhibited temperature spikes in November and December 2023 (see Fig.~\ref{fig:ni9}{b}). These anomalies may be attributed to maintenance activities conducted in these rooms.

\begin{figure}[!tb] 
\centering
\subfloat[]{\includegraphics[width=\columnwidth]{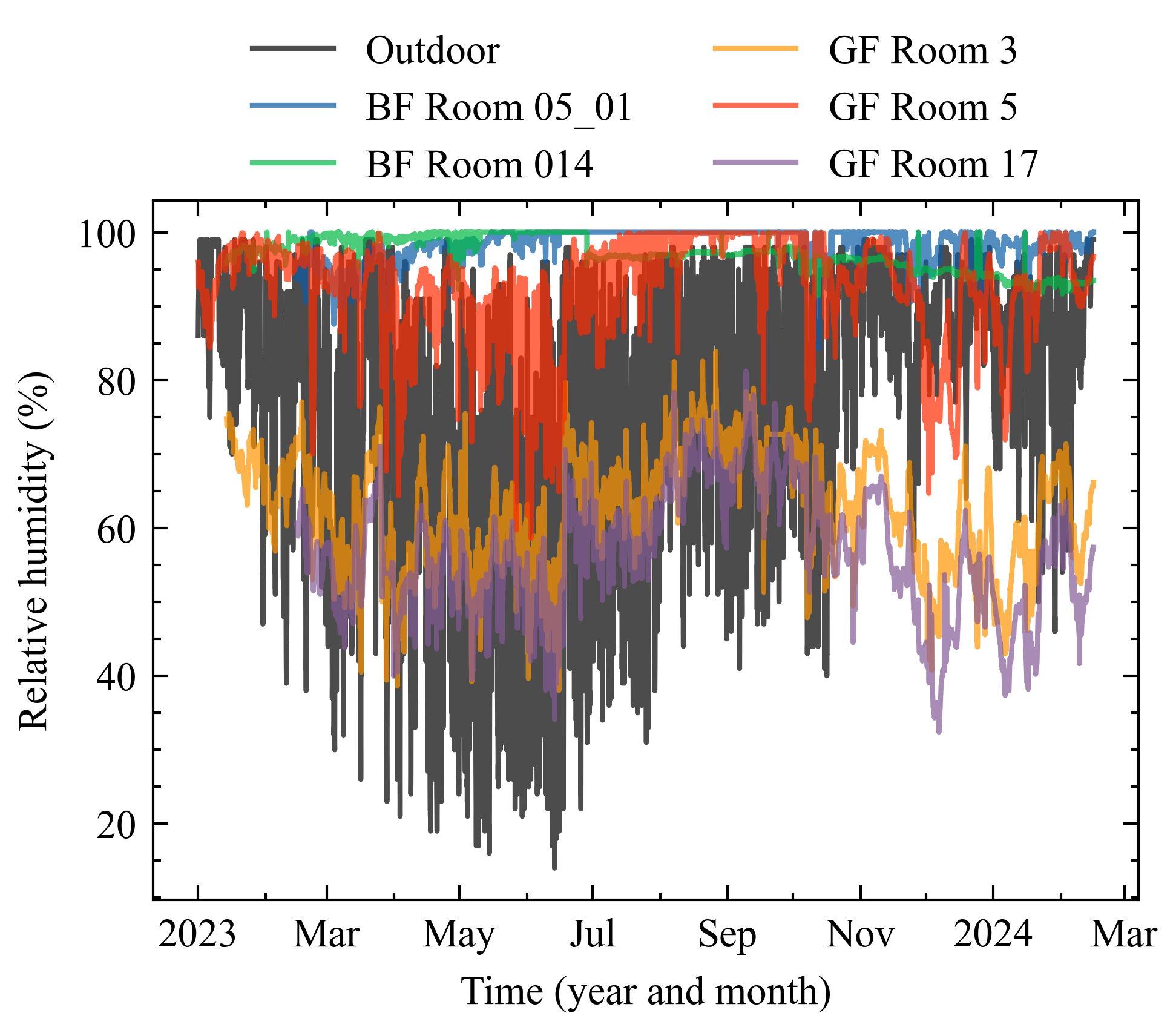}}
\hfill
\subfloat[]{\includegraphics[width=\columnwidth]{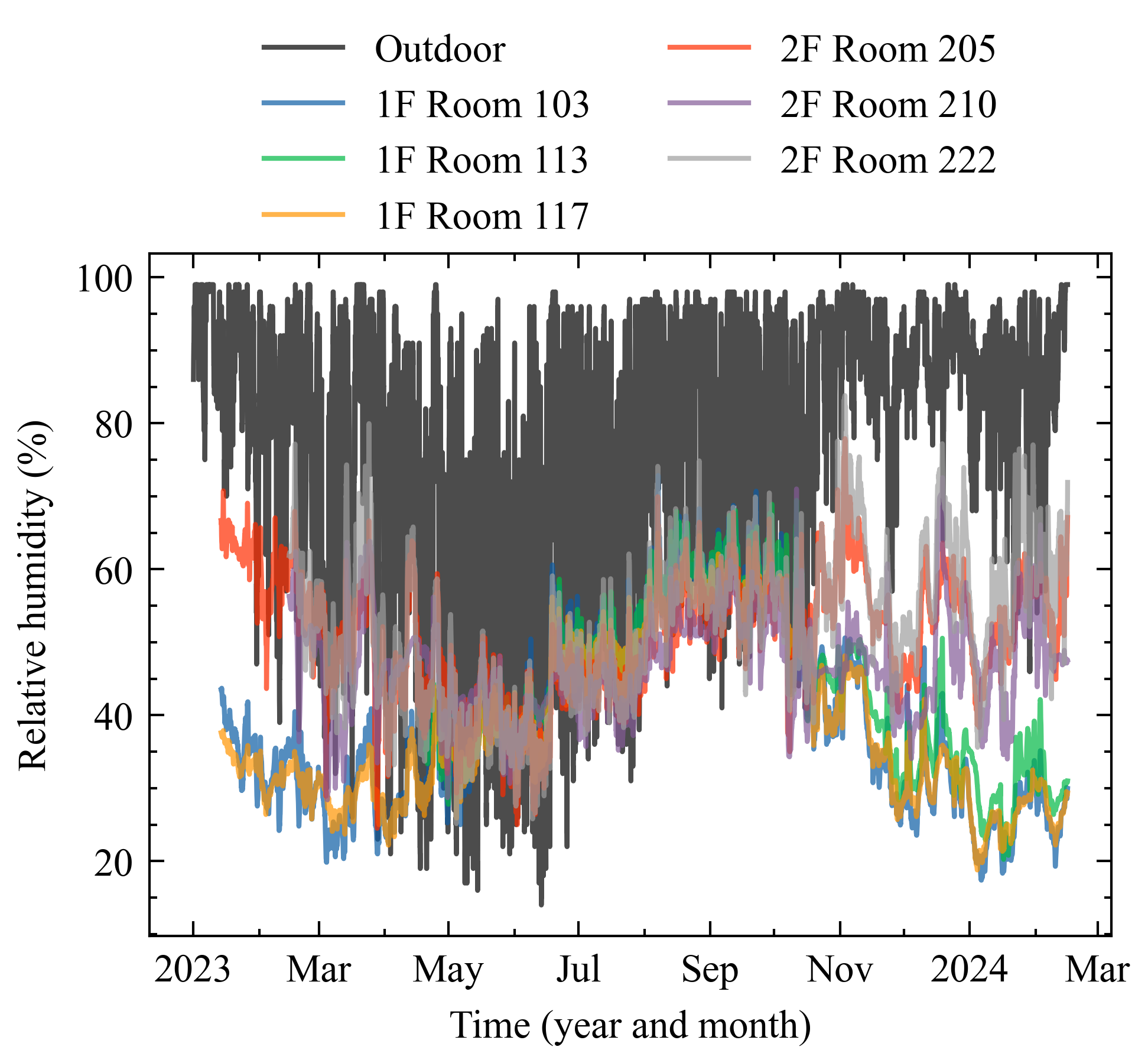}}
\caption{Historical hourly outdoor and indoor relative humidity of (\textbf{a}) rooms in the basement and on the ground floor as well as (\textbf{b}) rooms on the first and second floors from January 2023 to February 2024.}
\label{fig:ni10}
\end{figure}

Similarly, indoor RH tends to fluctuate less than outdoor RH. As shown in Fig.~\ref{fig:ni10}{a} and Fig.~\ref{fig:ni10}{b}, the difference between the annual maximum and minimum RH in most rooms was less than 40\%. In contrast, the difference in outdoor RH exceeded 80\%.

Rooms on the same floor typically have similar levels of RH. As depicted in Fig.~\ref{fig:ni10}{a}, RH in both basement rooms remained close to 100\% throughout the year. This extremely high humidity is primarily due to the direct soil flooring, which allows groundwater to continuously evaporate from the wet soil into the indoor air. Likewise, Room 3 and Room 17 on the ground floor (see Fig.~\ref{fig:ni10}{a}), as well as the three rooms on both the first and second floors (see Fig.~\ref{fig:ni10}{b}), each have approximate levels of RH. An exception was observed in Room 5 on the ground floor (see Fig.~\ref{fig:ni10}{a}), where the sensor box detected higher RH compared to other rooms on the same floor. The RH frequently exceeded 80\% and approached 100\% from July to October 2023. The high RH readings are likely caused by the evaporation of water from the ground and moisture from the basement rise through walls.

There are notable differences in RH across floors. As illustrated in Fig.~\ref{fig:ni10}{a} and Fig.~\ref{fig:ni10}{b}, in rooms without heating systems (BF, GF, and 2F), RH decreased as the floor goes up. Conversely, in rooms with heating systems (1F), RH was lower compared to rooms on the second floor when the heating system was working. The Mann-Whitney U test results also indicated a significant difference in indoor RH between rooms on different floors ($p < 0.001$).

\subsection{Comparison Between Indoor and Outdoor Humidity Mixing Ratios}

\begin{figure}[!tb] 
\centering
\subfloat[]{\includegraphics[width=\columnwidth]{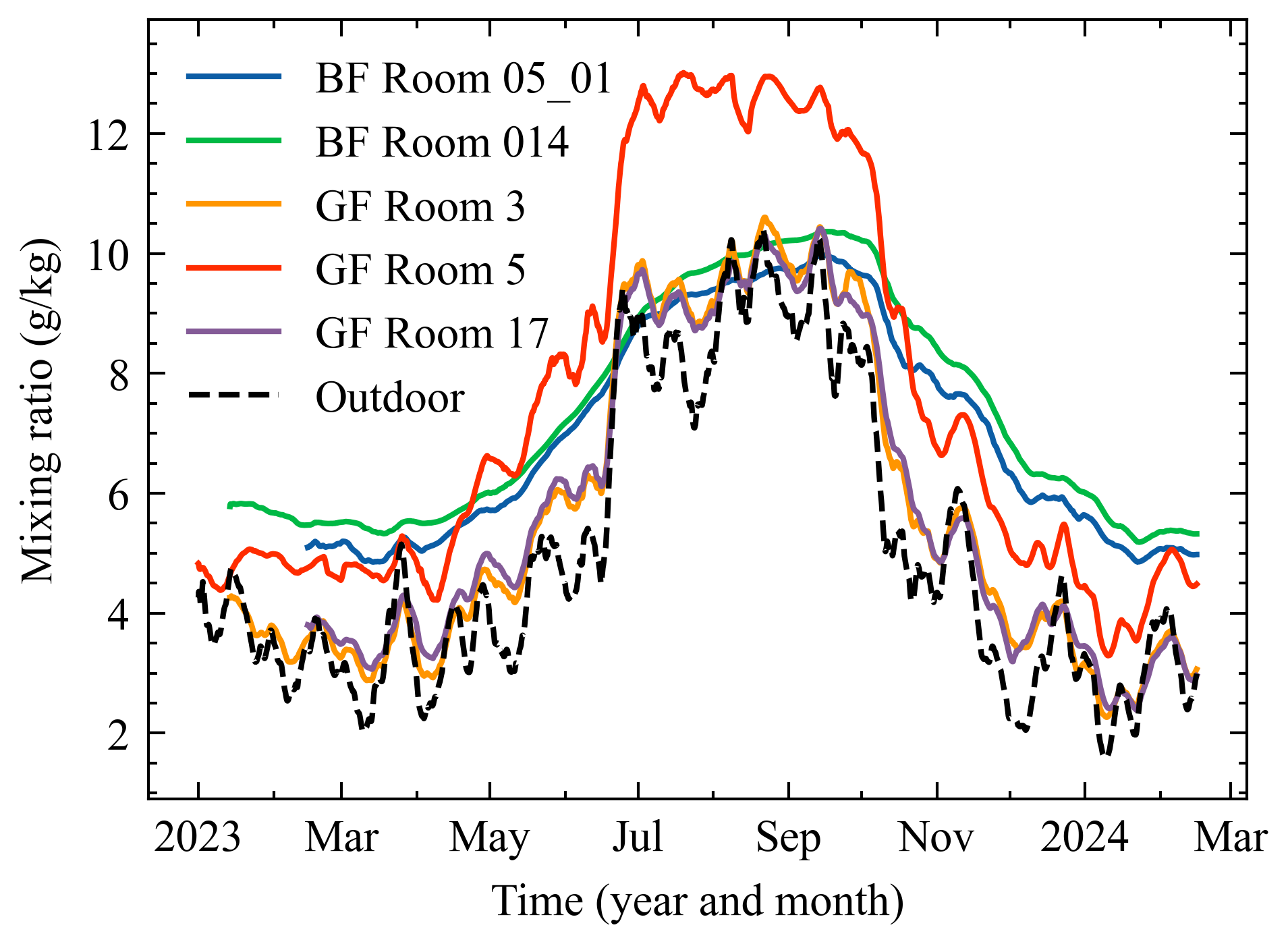}}
\hfill
\subfloat[]{\includegraphics[width=\columnwidth]{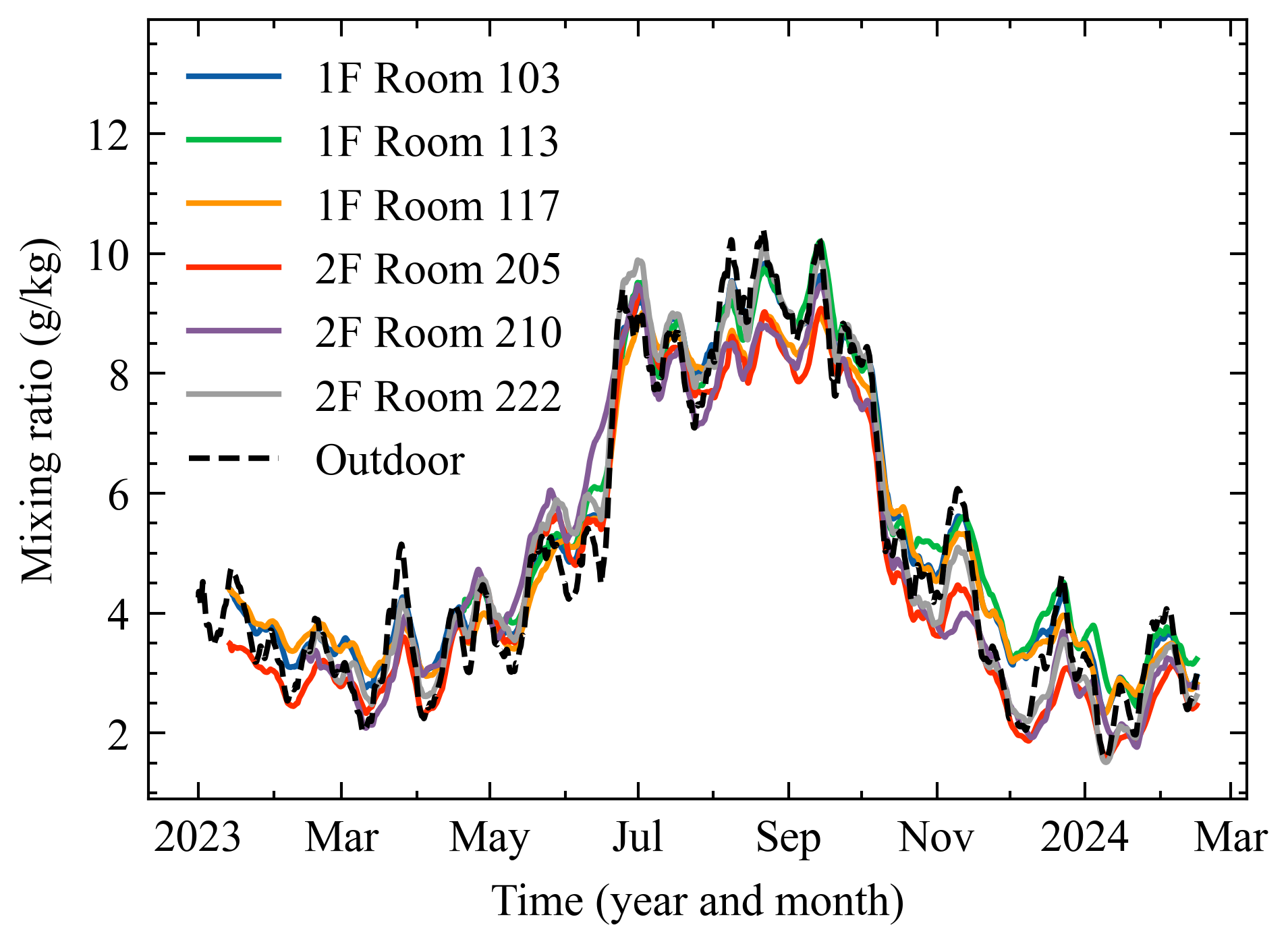}}
\caption{Historical 7-day moving average of hourly outdoor and indoor mixing ratios (MRs) of (\textbf{a}) rooms in the basement and on the ground floor as wells as (\textbf{b}) rooms on the first and second floors from January 2023 to February 2024.}
\label{fig:ni11}
\end{figure}

The previous analysis has shown that RH is excessively high in the basement and Room 5 on the ground floor. The high humidity in the basement is mainly due to the direct soil flooring, which allows groundwater continuously evaporates from the wet soil into the indoor air. This moisture is speculated to rise through the walls, acting as a moisture source that affects not only the basement but also rooms on the upper floors. As depicted in Fig.~\ref{fig:ni11}{a}, MRs in the basement and ground floor rooms were consistently higher than outdoors, suggesting an internal moisture source. The problem of high humidity is particularly pronounced in Room 5, where the sensor box is positioned on the floor and near the walls. The rising dampness has led to high humidity in Room 5. Thus, the evaporation from the basement need to be handled to take control of this problem. One measure is to install vapor barriers in the basement to reduce evaporation and moisture migration.

Rooms on the first and second floors are less affected by high RH in the basement. As shown in Fig.~\ref{fig:ni11}{b}, the difference between MRs of these rooms and outdoors was smaller than that of the basement and the ground floor. Furthermore, MRs of rooms on the second floor were equal to or less than outdoors for most of the year.

\begin{figure}[!tb]
\includegraphics[width=\columnwidth]{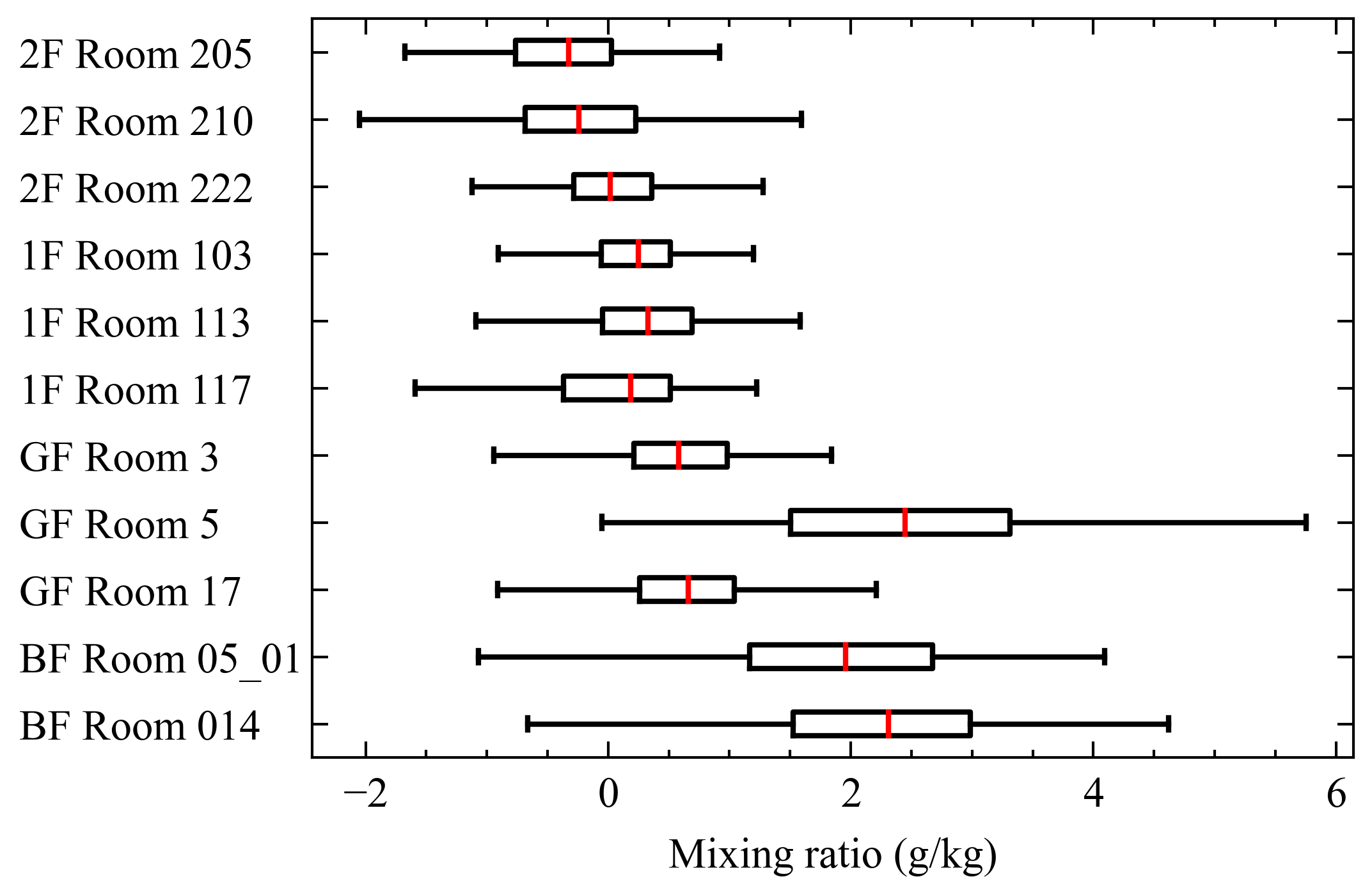}
\centering
\caption{Boxplot of the difference between indoor and outdoor MRs from January 2023 to February 2024. Each difference is calculated using a 7-day moving average of hourly indoor MR minus a 7-day moving average of hourly outdoor MR. Outliers are not shown since they are not of interest in this study.}\label{fig:ni12}
\end{figure}

The impact of high humidity in the basement on upper floors is more clearly illustrated in Fig.~\ref{fig:ni12}. This figure presents a boxplot of the difference between indoor and outdoor MRs from January 2023 to February 2024. Each difference was calculated using a 7-day moving average of hourly indoor MRs minus a 7-day moving average of hourly outdoor MRs. Outliers have been excluded as they are not relevant to this study. As shown in Fig.~\ref{fig:ni12}, the basement and ground floor rooms consistently exhibited higher MRs compared to the outdoor environment for approximately 75\% of the observed period. This trend gradually decreased on upper floors of the building. On the second floor, the indoor MRs exceeded outdoor levels roughly 50\% of the time, indicating a more balanced relationship where the indoor ratios are lower than outdoor levels for the other half of the time. These findings strongly suggest the presence of a moisture source originating from the basement.

\subsection{Mold Risk}

\begin{figure}[!tb]
\includegraphics[width=\columnwidth]{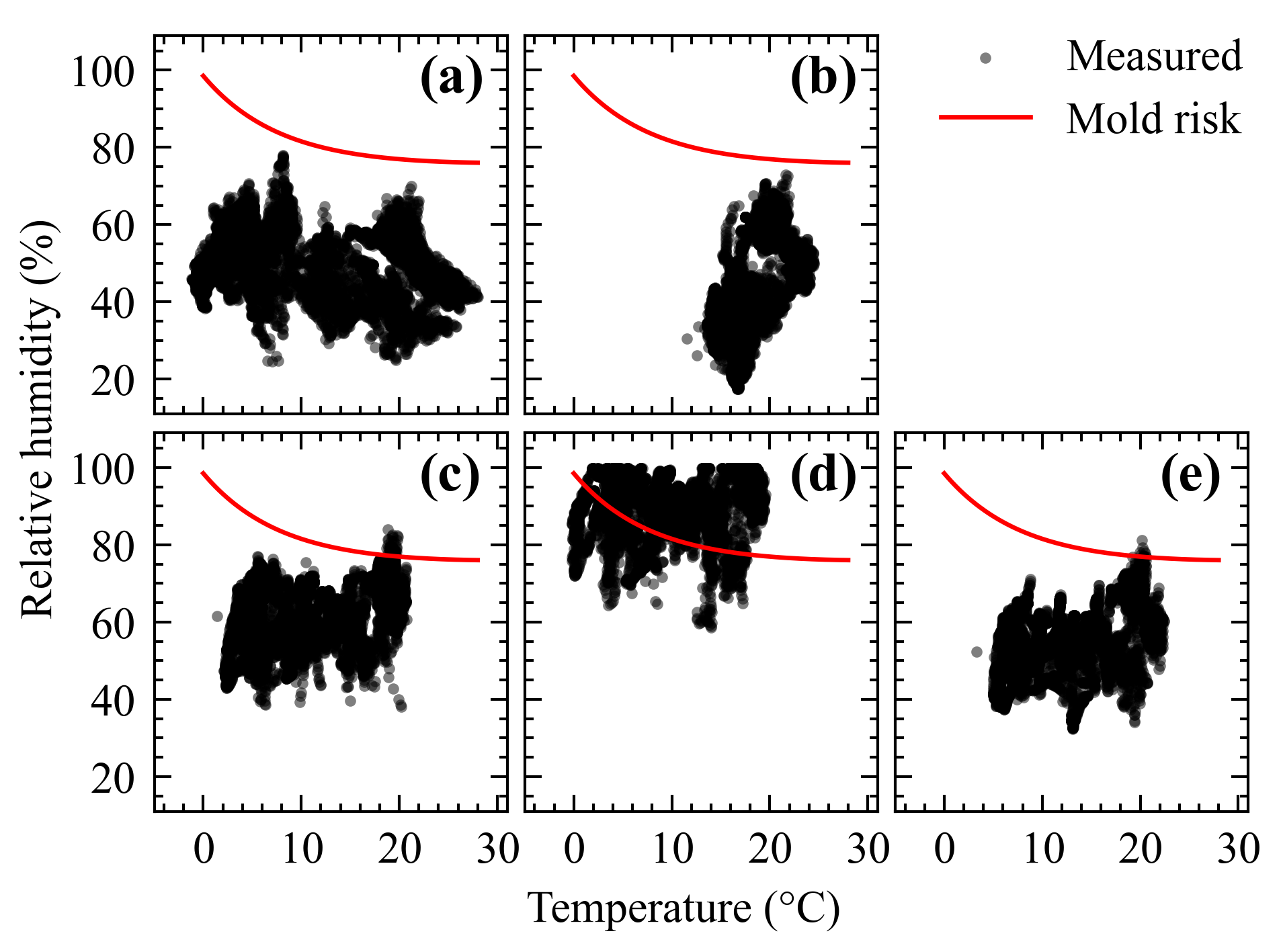}
\centering
\caption{Mold risk in (\textbf{a}) Room 205, (\textbf{b}) Room 103, (\textbf{c}) Room 3, (\textbf{d}) Room 5, and (\textbf{e}) Room 17 from January 2023 to February 2024. Measurements above the red curve are identified as risky points.}\label{fig:ni13}
\end{figure}

High humidity levels can lead to problems such as mold growth. As shown in Fig.~\ref{fig:ni13}{c}--Fig.~\ref{fig:ni13}{e}, varying degrees of mold risk were observed in three rooms on the ground floor. Among these, Room 5 had the highest mold risk, followed by Room 3, with Room 17 showing the lowest risk. This variation may be due to the different installation heights of the sensor boxes, with Room 17 having the highest placement and Room 5 the lowest. In contrast, no mold risk was detected in selected rooms on the first and second floors (see Fig.~\ref{fig:ni13}{a} and Fig.~\ref{fig:ni13}{b}). This finding further confirms that continuous evaporation in the basement primarily affects the basement and ground floor, with negligible impact on the first and second floors.

\subsection{Short-term Fluctuation of Relative Humidity}

\begin{figure}[!tb]
\includegraphics[width=\columnwidth]{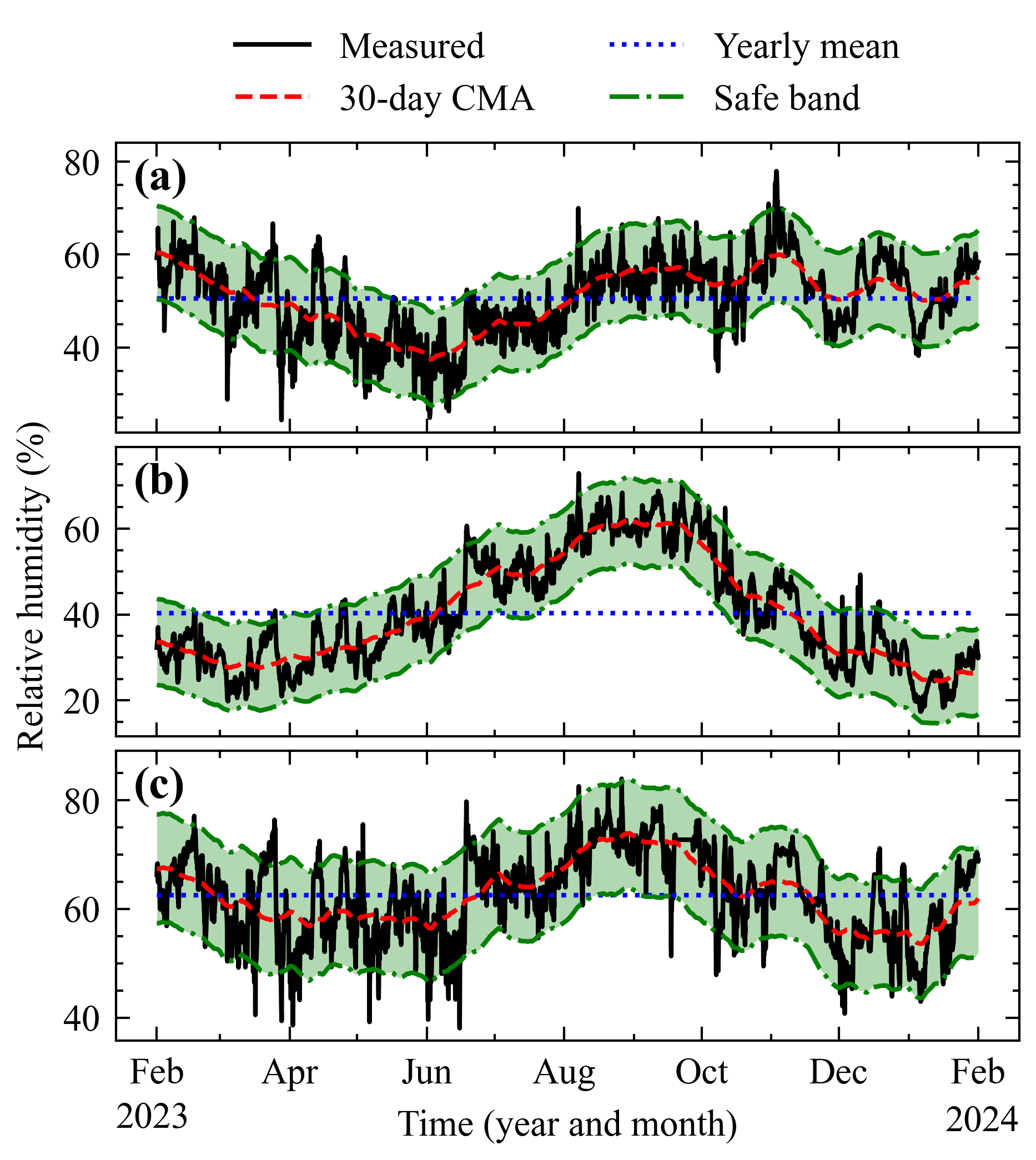}
\centering
\caption{A full calendar year of historical relative humidity (RH) measured in (\textbf{a}) Room 205, (\textbf{b}) Room 103, and (\textbf{c}) Room 3 from February 1, 2023 to January 31, 2024. A seasonal RH cycle is obtained by calculating the 30-day centered moving average (CMA) of RH readings. Safe band is determined by 10\% apart from the 30-day CMA. Measurements outside the safe band are identified as risky points.}\label{fig:ni14}
\end{figure}

Aside from the basement, rooms on lower floors exhibited greater short-term RH fluctuations. As illustrated in Fig.~\ref{fig:ni14}{c}, Room 3 was identified as particularly at risk. Room 103 (see Fig.~\ref{fig:ni14}{b}) experienced the fewest risky periods when the heating system was working. In comparison, Room 205 (see Fig.~\ref{fig:ni14}{a}) had fewer risky periods than Room 3 but more than Room 103, indicating it was less affected by water evaporation from the basement.

Since rooms on the first floor maintain relatively stable RH levels, continuous heating throughout the winter might be unnecessary. Implementing more energy-efficient heating strategies, such as intermittent heating, can ensure human comfort during occupancy—especially during long gatherings in Room 103 with large groups—while also conserving energy. Additionally, a zoned heating approach could allow for targeted heating in different areas of the building based on conservation needs and occupancy patterns. This strategy not only conserves energy but also improves occupant comfort by providing customized climate control in frequently used spaces while maintaining lower temperatures in less occupied areas.

\subsection{Impact of Precipitation on Groundwater Level Changes}

\begin{figure}[!tb]
\includegraphics[width=\columnwidth]{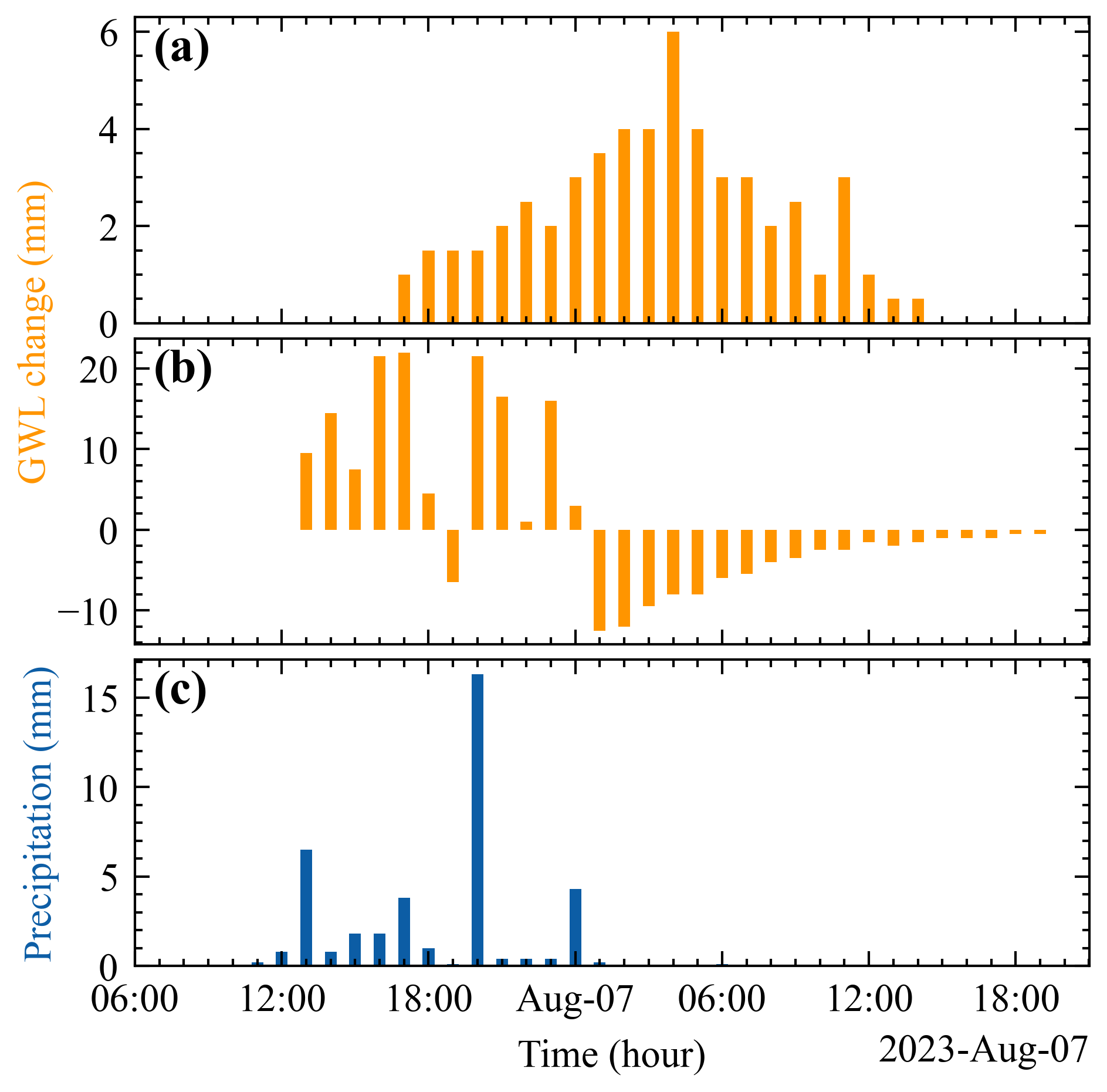}
\centering
\caption{Hourly GWL changes in (\textbf{a}) Ch.1 and (\textbf{b}) Ch.2 of Room 014, along with (\textbf{c}) hourly precipitation, from 8:00 August 6 to 20:00 August 7, 2023.}\label{fig:ni15}
\end{figure}

In the basement, the exposed soil of the floor contains abundant groundwater, which is evidently influenced by precipitation. Fig.~\ref{fig:ni15} illustrates the effect of hourly outdoor precipitation on the hourly changes in GWL at two monitoring points in Room 014 of the basement. Channel 2, located on the east side, and Channel 1, on the west side, showed different responses to rainfall.

As shown in Fig.~\ref{fig:ni15}{b}, the GWL at Channel 2 responded almost immediately to rainfall (see Fig.~\ref{fig:ni15}{c}), with virtually no delay. After the precipitation ended, the GWL at Channel 2 gradually decreased and eventually stabilized. In contrast, the GWL at Channel 1 exhibited a delay of about five hours (see Fig.~\ref{fig:ni15}{a}). This delay is likely due to precipitation infiltrating the ground from the courtyard on the east side of the main building and gradually spreading from east to west.

Changes in the GWL at Channel 2 were more pronounced than those at Channel 1. As illustrated in Fig.~\ref{fig:ni15}{a} and Fig.~\ref{fig:ni15}{b}, the GWL at Channel 2 can vary by up to 20~mm per hour, while at Channel 1, the variation was up to 6~mm per hour.

\begin{figure}[!tb]
\includegraphics[width=\columnwidth]{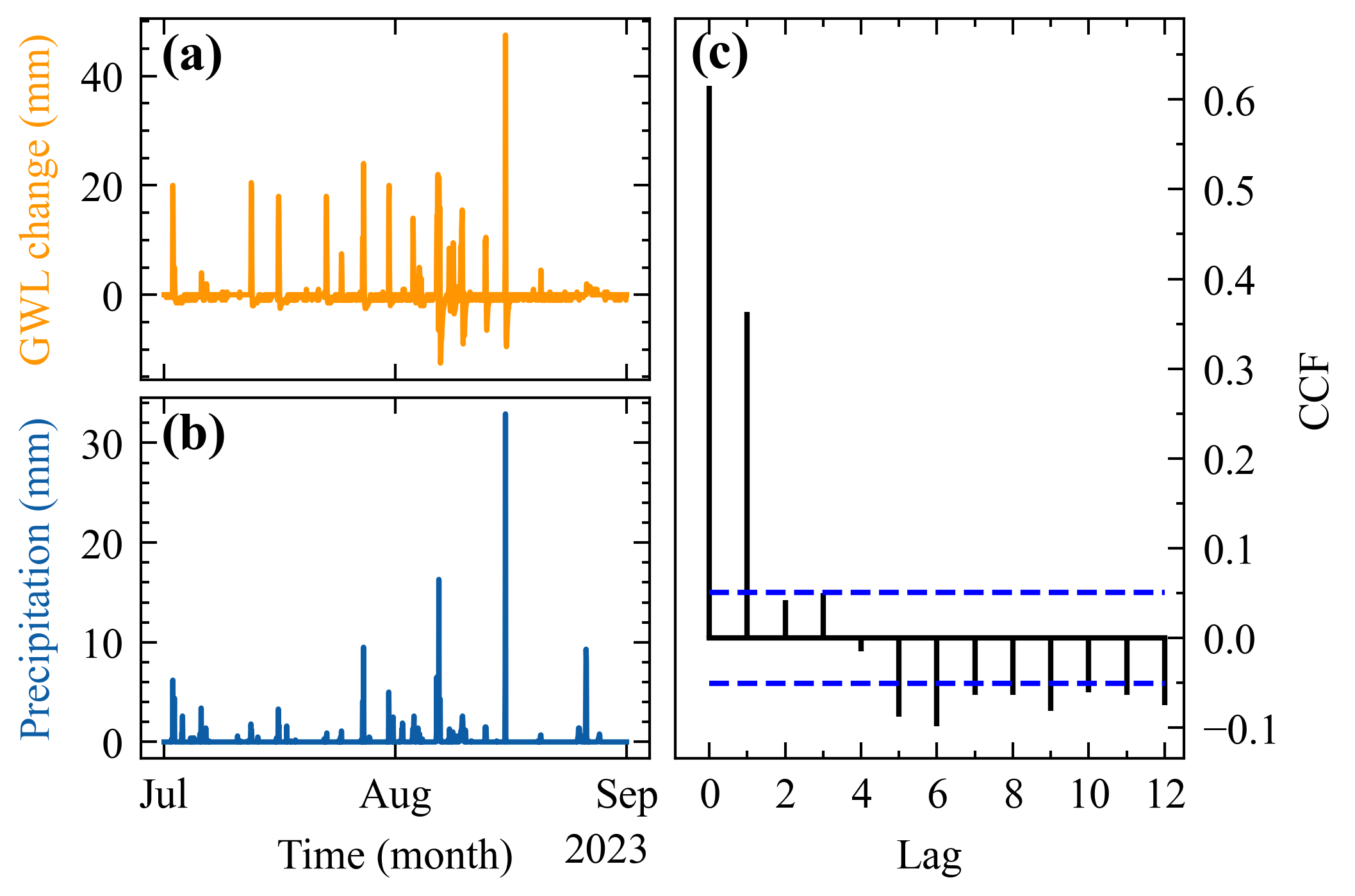}
\centering
\caption{Hourly GWL changes in (\textbf{a}) Ch.2 of Room 014,  (\textbf{b}) hourly precipitation from July 1 to August 31, 2023, and (\textbf{c}) their cross-correlation function (CCF) values.}\label{fig:ni16}
\end{figure}

An analysis of the data for July and August in 2023 confirms a correlation between changes in GWL and precipitation. The cross-correlation analysis (see Fig.~\ref{fig:ni16}) reveals that the maximum cross-correlation function (CCF) value is 0.62$\pm$0.05 at lag$=$0 with a 95\% confidence interval. This finding indicates that the GWL at Channel 2 responds to rainfall almost immediately, with negligible delay.

In addition to installing vapor barriers in the basement to mitigate evaporation, another measure can be constructing an underground exterior wall on the east side of the main building to prevent rainwater from seeping into the basement.

\subsection{Impact of Occupants on Indoor Climate}

\begin{figure}[!tb]
\includegraphics[width=\columnwidth]{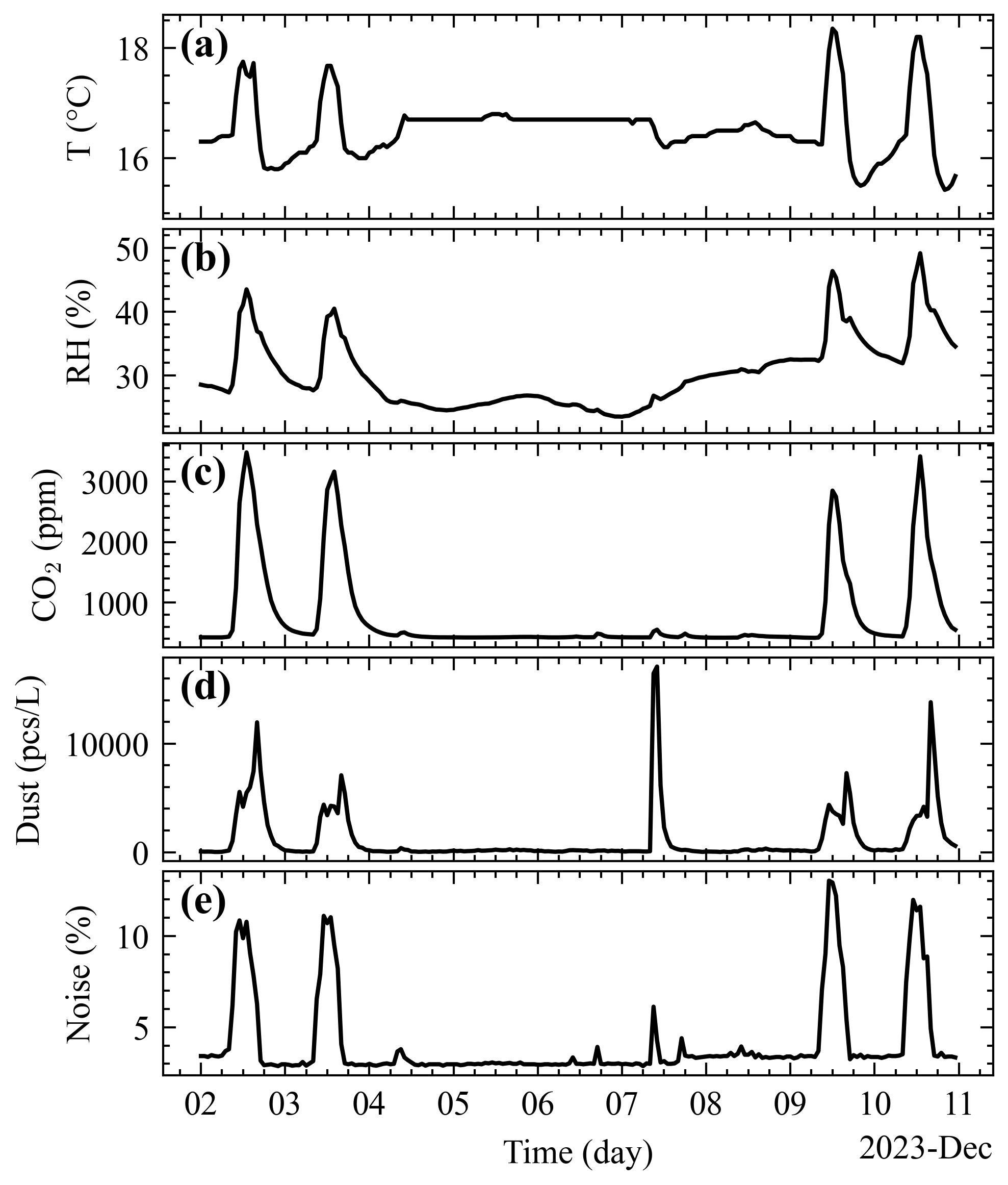}
\centering
\caption{Historical hourly indoor climate of Room 103 from December 2 to 11, 2023. (\textbf{a}) Temperature (T), (\textbf{b}) RH, (\textbf{c}) CO\textsubscript{2} concentration, (\textbf{d}) dust concentration, and (\textbf{e}) noise relative intensity.}\label{fig:ni17}
\end{figure}

\begin{figure}[!tb]
\includegraphics[width=\columnwidth]{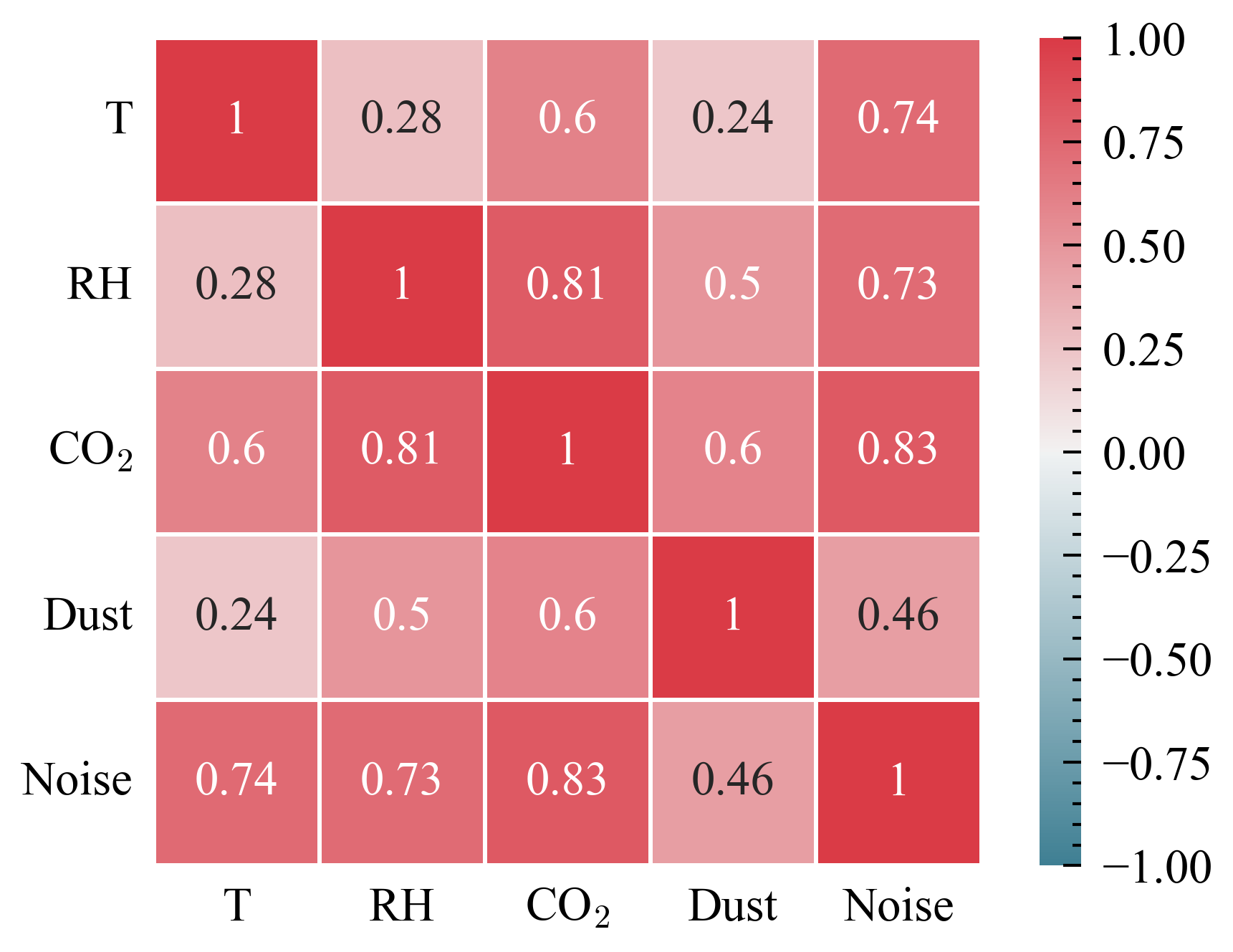}
\centering
\caption{The matrix of Pearson correlation coefficients between five indoor environmental parameters of Room 103 from December 2 to 11, 2023.}\label{fig:ni18}
\end{figure}

Löfstad Castle is open for public activities. The presence of occupants can impact indoor climate. Fig.~\ref{fig:ni17} illustrates changes in five indoor environmental parameters—temperature, RH, CO\textsubscript{2} concentration, dust concentration, and noise level—in Room 103 over nine days from December 2 to 10, 2023. Visitor gatherings occurred during the daytime on December 2, 3, 9, and 10. During these gatherings, all five environmental parameters increased considerably. Pearson correlation analysis confirmed the impact of occupant presence on environmental parameters. As shown in Fig.~\ref{fig:ni18}, CO\textsubscript{2} concentration had a strong positive correlation with temperature ($r = 0.6$) and dust concentration ($r = 0.6$), and a very strong positive correlation with RH ($r = 0.81$) and noise level ($r = 0.83$). The highest CO\textsubscript{2} concentration on these days even reached or exceeded 3000 ppm, which could cause discomfort for occupants~\cite{ansi_standard_62}. Besides, there seems to have maintenance work on December 7, as indicated by a significant rise in dust concentration and a slight increase in noise levels (see Fig.~\ref{fig:ni17}{d} and Fig.~\ref{fig:ni17}{e}).

Due to natural ventilation, it takes time for indoor environment to return to its regular conditions after such events. For instance, CO\textsubscript{2} concentration required over six hours to drop back to pre-event levels (see Fig.~\ref{fig:ni17}{c}), while dust concentration decreased more quickly (see Fig.~\ref{fig:ni17}{d}). As shown in Fig.~\ref{fig:ni17}{a}, temperature can also return to the preset level promptly, thanks to the regulation of the heating system. In contrast, RH recovered more slowly than temperature (see Fig.~\ref{fig:ni17}{b}). 

\section{Conclusion}
\label{sec:conclusion}
This study aimed to create parametric digital twins to preserve historic buildings. Parametric digital twins were developed by integrating various digital technologies such as Internet of Things, edge and cloud computing, and ontology. A comprehensive case study was conducted at Löfstad Castle, a listed historic building in Östergötland, Sweden. Thirteen cloud-connected sensor boxes with a total of 84 sensors were installed throughout the main building to facilitate the creation of the digital twin. 

The results revealed the presence of a moisture source originating from the basement. This moisture source also affected rooms on upper floors, particularly the ground floor, and posed risks for mold growth and material deterioration caused by strain-stress cycles. It was also discovered that current natural ventilation could not adequately manage indoor air quality during large and prolonged gatherings. The findings suggested taking measures like installing vapor barriers to address humidity problems in the basement and on the ground floor, thus mitigating risks such as mold growth and material deterioration. Additionally, energy-efficient heating strategies were suggested for upper floors. 

This study contributes to the field of historic building conservation by demonstrating a comprehensive digitalization solution and its application. The scalable and adaptable solution supports diverse historic buildings beyond Löfstad Castle. Its modular design allows the integration of additional sensors and data sources, making it suitable for diverse architectural layouts and conservation needs. Using cloud-based storage and analysis, the system can handle large datasets from complex buildings, enabling data-driven conservation. Furthermore, adaptability of the solution allows customization for specific environmental monitoring needs and heritage contexts. It provides conservation practitioners with applicable tools for tracking environmental conditions, analyzing long-term trends, and implementing sustainable preservation measures. The solution and findings from this study thus offer insights and practical guidance for the conservation of other historic buildings facing similar challenges.

\section*{Acknowledgment}

The authors thank Gustav Knutsson at Linköping University for helping prepare materials, produce sensor boxes, and install devices. Per Stålebro, Emma Vilhelmsson, and Morgan Brålin at Östergötland's Museum are acknowledged for providing access to Löfstad Castle and facilitating the installation of devices. Aurora Ekman and Helena Sandekull at Fredriksson Arkitektkontor AB provided the floor plan and section drawing of Löfstad Castle. The Swedish Meteorological and Hydrological Institute in Norrköping supplied application programming interfaces for accessing meteorological data.

\bibliographystyle{IEEEtran}
\bibliography{references}

\begin{IEEEbiography}[{\includegraphics[width=1in,height=1.25in,clip,keepaspectratio]{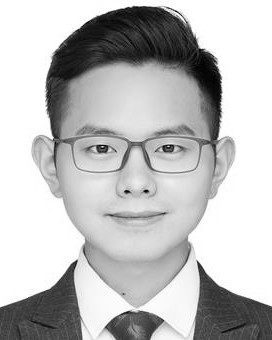}}]{Zhongjun Ni} received the B.Eng. and M.Eng. degrees from Zhejiang University, China, in 2014 and 2017, respectively, and the Licentiate of Engineering degree from Linköping University, Sweden, in 2023. Between 2017 and 2020, he worked as a software engineer in the industry, such as Microsoft. He is currently pursuing his Ph.D. degree with the Department of Science and Technology at Linköping University, Sweden. His research interests include time series analysis, digital twins, and Internet of Things solutions based on Edge-Cloud computing.
\end{IEEEbiography}

\begin{IEEEbiography}[{\includegraphics[width=1in,height=1.25in,clip,keepaspectratio]{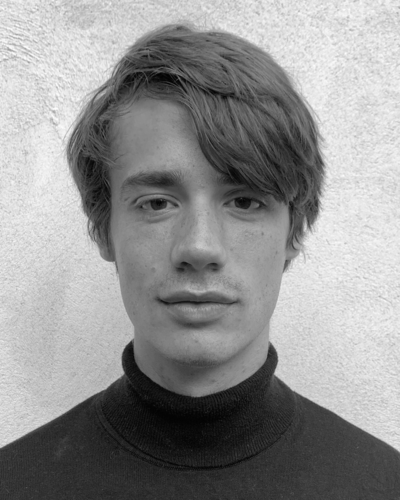}}]{Jelrik Hupkes} received the B.A. degree in art history from Ghent University, Belgium, in 2021 and the M.A. degree in conservation with a focus on cultural heritage studies from Uppsala University, Sweden, in 2023. He also received the B.F.A. degree in film from RITCS School of Arts, Belgium, in 2019. Between November 2023 and June 2024, he worked as a research assistant with the Department of Art History, Conservation at Uppsala University. His research interest is mainly focused on critical perspectives on architectural heritage from the 19th and 20th centuries as well as the relationship between architecture and film.     
\end{IEEEbiography}

\begin{IEEEbiography}[{\includegraphics[width=1in,height=1.25in,clip,keepaspectratio]{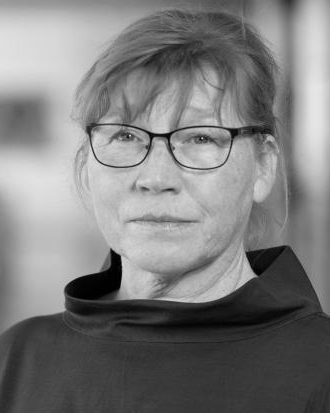}}]{Petra Eriksson} received the Ph.D. degree in conservation from the University of Gothenburg, Sweden, in 2021. She has worked as a teacher and researcher in conservation with the Department of Art History, Conservation at Uppsala University since 2013 and with University of Gotland since 2003. Her research focuses on heritage values in buildings and energy efficiency in historic buildings.
\end{IEEEbiography}

\begin{IEEEbiography}[{\includegraphics[width=1in,height=1.25in,clip,keepaspectratio]{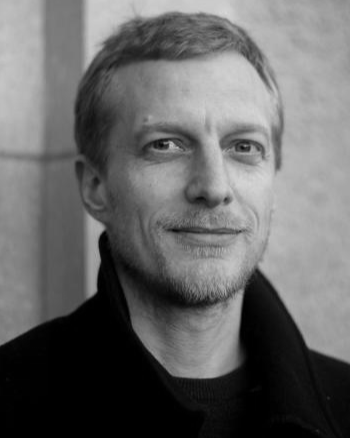}}]{Gustaf Leijonhufvud} received the Ph.D. degree in conservation from the University of Gothenburg, Sweden, in 2016. Currently, he works as a Senior Lecturer with the Department of Art History, Conservation at Uppsala University. His research interest is about sustainable management of cultural heritage, with a focus on energy and indoor climate issues. He is especially interested in the interplay between research, policy, and practice in this field.
\end{IEEEbiography}

\begin{IEEEbiography}[{\includegraphics[width=1in,height=1.25in,clip,keepaspectratio]{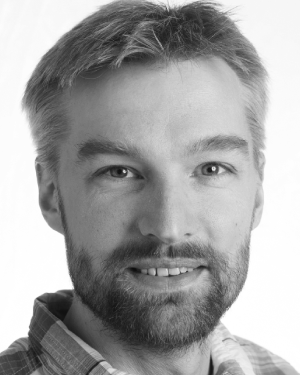}}]{Magnus Karlsson} received the M.Sc., Licentiate of Engineering, and Ph.D. degrees from Linköping University, Sweden, in 2002, 2005, and 2008, respectively. In 2003, he joined the Communication Electronics Research Group, Linköping University, where he is currently a Senior Associate Professor. Apart from his broad interest in electronic system design and microwave technology in particular, his main work involves wideband transceiver and antenna techniques, and wireless communication. The later includes high speed data transmission and sensor networks and its associated applications.
\end{IEEEbiography}

\begin{IEEEbiography}[{\includegraphics[width=1in,height=1.25in,clip,keepaspectratio]{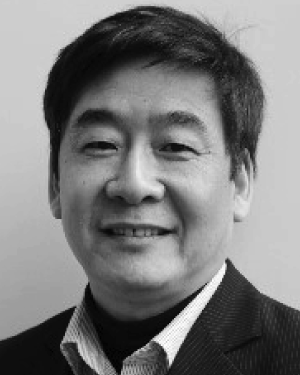}}]{Shaofang Gong} received the B.Sc. degree in microelectronics from Fudan University in Shanghai, China, in 1982, as well as the Licentiate of Engineering and Ph.D. degrees from Linköping University, Sweden, in 1988 and 1990, respectively. Between 1991 and 1999, he was a Senior Researcher with the research institute RISE Acreo, Sweden. From 2000 to 2001, he was the Chief Technology Officer with a spin-off company from the research institute. In the mean time, he was an Adjunct Professor with Linköping University. Since 2002, he has been the Chair Professor of Communication Electronics with Linköping University. His main research interests include communication electronics including radio frequency and microwave system design, high speed data transmissions, and wireless sensor networks towards Internet of Things.
\end{IEEEbiography}

\EOD

\end{document}